\documentclass[sigconf]{acmart}

\usepackage{siunitx}
\usepackage[flushleft]{threeparttable}

\begin{document}

\settopmatter{printacmref=false} 

\onecolumn

\section*{IEEE Copyright Notice}

\textcopyright$\;$2020 IEEE. Personal use of this material is permitted. Permission from IEEE must be obtained for all other uses, in any current or future media, including reprinting/republishing this material for advertising or promotional purposes, creating new collective works, for resale or redistribution to servers or lists, or reuse of any copyrighted component of this work in other works. 

\textbf{Accepted to be published in: Proceedings of the 2020 International Conference On Computer Aided Design (ICCAD), Nov 2-5, 2020, San Diego, CA, USA.}
\twocolumn
\newpage

\title{BoMaNet: Boolean Masking of an Entire Neural Network}

\author{Anuj Dubey}
\email{aanujdu@ncsu.edu}
\affiliation{
  \institution{North Carolina State University}
  \city{Raleigh}
  \state{North Carolina}
  \postcode{27606}
}

\author{Rosario Cammarota}
\email{rosario.cammarota@intel.com}
\affiliation{
  \institution{Intel Labs}
  \city{San Diego}
  \country{United States}}

\author{Aydin Aysu}
\email{aaysu@ncsu.edu}
\affiliation{
  \institution{North Carolina State University}
  \city{Raleigh}
  \state{North Carolina} 
  \postcode{27606}
}

\begin{abstract}
Recent work on stealing machine learning (ML) models from inference engines with physical side-channel attacks warrant an urgent need for effective side-channel defenses. 
This work proposes the first \emph{fully-masked} neural network inference engine design. 

Masking uses secure multi-party computation to split the secrets into random shares and to decorrelate the statistical relation of secret-dependent computations to side-channels (e.g., the power draw). 
In this work, we construct secure hardware primitives to mask \emph{all} the linear and non-linear operations in a neural network. 
We address the challenge of masking integer addition by converting each addition into a sequence of XOR and AND gates and by augmenting Trichina's secure Boolean masking style. We improve the traditional Trichina's AND gates by adding pipelining elements for better glitch-resistance and we architect the whole design to sustain a throughput of 1 masked addition per cycle. 

We implement the proposed secure inference engine on a Xilinx Spartan-6 (XC6SLX75) FPGA. The results show that masking incurs an overhead of 3.5\% in latency and 5.9$\times$ in area. 
Finally, we demonstrate the security of the masked design with 2M traces. 
\vspace{-0.75em}
\end{abstract}

\keywords{\vspace{-0.4em}Masking, neural networks, side-channel attacks, model stealing}

\maketitle

\section{Introduction}
\vspace{-0.25em}
Physical side-channel attacks pose a major threat to the security of cryptographic devices. Attacks like the Differential Power Analysis (DPA) \cite{C:KocJafJun99} can extract secret keys by exploiting the inherent correlation between the secret-key-dependent data being processed and the Complementary Metal Oxide Semiconductor (CMOS) power consumption \cite{das2019stellar}. DPA has been shown to be effective against many cryptographic implementations in the last two decades 
\cite{mangard2008power,TCHES:PSKH18,chen2015differential}.
Until recently, these attacks were confined to cryptographic schemes.
But lately, the Machine Learning (ML) applications are shown to be vulnerable to physical side-channel attacks~\cite{batina2018csi,dubey2019maskednet,yudeepem}, where an adversary aims to reverse engineer the ML model. 
Indeed, these models are lucrative targets as they are costly to develop and hence become valuable IPs for the companies \cite{strubell2019energy}. 
Knowledge about model parameters also makes it easy to fool the model using adversarial learning, which is a serious problem if the model performs a critical task like fraud/spam detection \cite{advlearn05}. 

Unfortunately, most of the existing work on the physical side-channel analysis of ML accelerators has focused only on attacks, not defenses. To date, there are three publications focusing specifically on the power/EM side-channel leakage of ML models. The first two discuss some countermeasures like shuffling and masking but do not implement any~\cite{batina2018csi,yudeepem}. 
The third one implements a hybrid of masking and hiding based countermeasures and exposes the vulnerability in the arithmetic masking of integers due to the leakage in the sign-bit~\cite{dubey2019maskednet}. 

Masking uses secure multi-party computation to split the secrets into random shares and to decorrelate the statistical relation of secret-dependent computations to side-channels.
Although similar work on cryptographic hardware has been fully masked \cite{BGK05}, the earlier work on neural network hardware was \emph{partially masked} for cost-effectiveness~\cite{dubey2019maskednet} while the leakage in the sign bit is hidden. Such solutions may work well for a regular IP where reasonable security at low-cost is sufficient. However, \emph{full masking} is a better alternative for IPs deployed in critical applications (like defense) requiring stronger defenses against side-channel attacks. 

In this work, we propose the design of the \emph{first fully-masked} neural network accelerator resistant against power-based side-channel attacks. 
We construct novel masked primitives for the various linear and non-linear operations in a neural network using gate-level \emph{Boolean} masking and masked look-up tables (LUT). We analyze neural network-specific computations like weighted summations in fully-connected layers and build specialized masked adders and multiplexers to perform those operations in a secure way. We also design a novel hardware that finds the greatest integer out of a set of integers in a masked fashion, which is needed in the output layer to find the node with the highest confidence score.

We target an area-optimized Binarized Neural Network (BNN) in our work because of their preference for edge-based neural network inference \cite{umuroglu2017finn,rastegari2016xnor}. We optimize the hardware design to reduce the impact of masking on the performance. Specifically, we build an innovative adder-accumulator architecture that provides a throughput of one addition per cycle even with a multi-cycle masked adder with feedback. We maximize the number of balanced data-paths in the masking elements by adding registers at every stage, to synchronize the arrival of signals and reduce the effects of glitches \cite{CHES:ManPraOsw05}. We build the masked design in a modular fashion starting from smaller blocks like Trichina's AND gates to finally build larger structures like the 20-bit masked full adder. We have pipelined the full design to maintain a high throughput.

Finally, we implement both the baseline unmasked and the proposed first-order secure masked neural network design on an FPGA. We use the standard TVLA methodology \cite{becker2013test} to evaluate the first-order security of the design and demonstrate no leakage up to 2M traces. The latency of the masked implementation is merely 3.5\% higher than the latency of the unmasked implementation. The area of the masked design is 5.9$\times$ that of the unmasked design. Our goal in this paper is to provide the first fully-masked design where we propose certain optimizations and a practical evaluation of security. We also discuss potential further optimizations and extensions of masking for hardware design and security refinements. 
\vspace{-2em}
\section{Threat Model}
\vspace{-0.25em}
We adopt the standard DPA threat model in which an adversary has direct physical access to the target device running inference~\cite{batina2018csi,dubey2019maskednet,kocher2011introduction}, or can obtain power measurements remotely~\cite{zhao2018fpga} when the device executes neural network computations. The adversary can control the inputs and observe the corresponding outputs from the device as in chosen-plaintext attacks. 
\begin{figure}
    \includegraphics[width=0.45\textwidth]{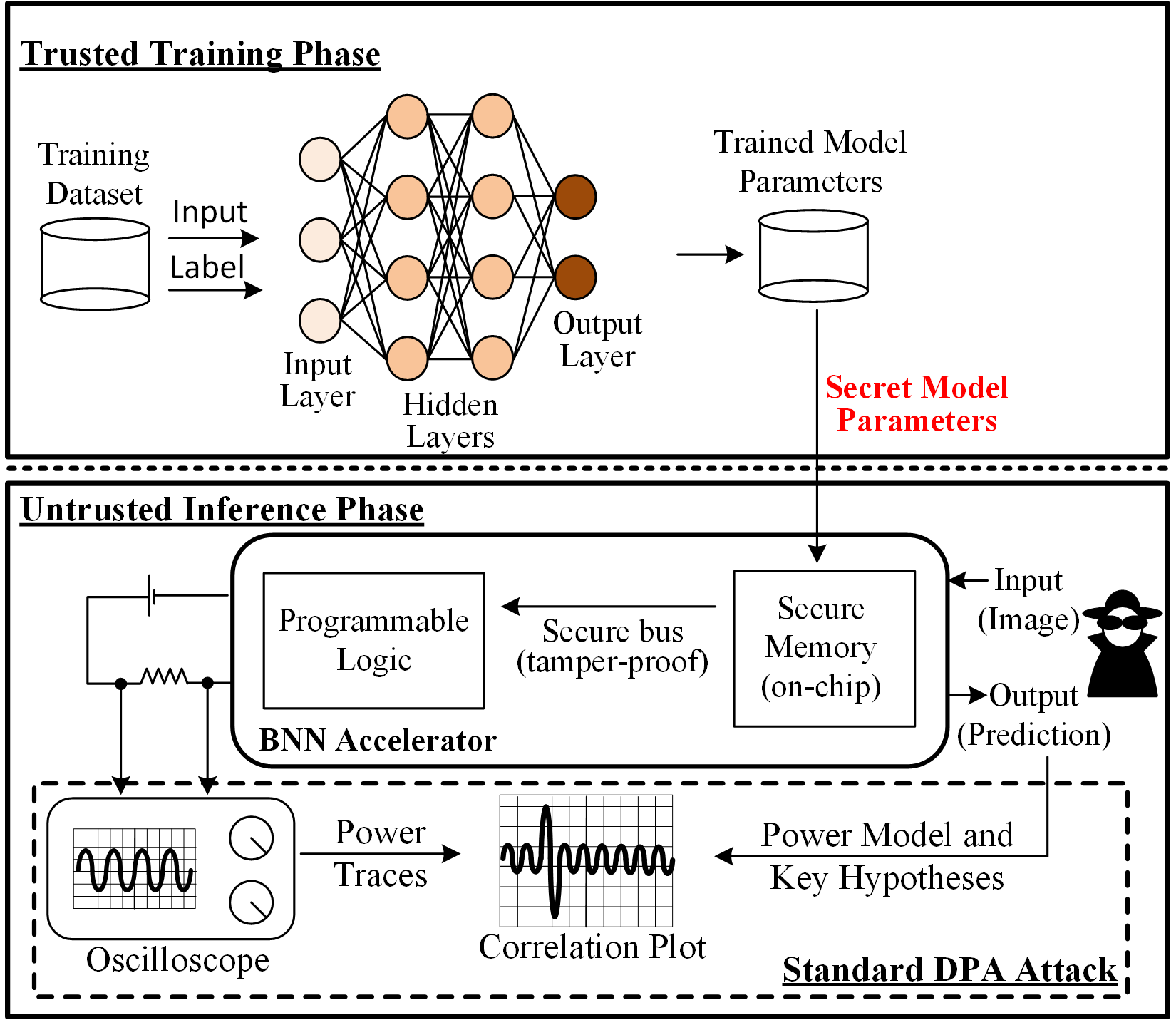}
    \vspace{-1.1em}
    \caption{Standard DPA threat model applied to ML model stealing, where the trained neural network is deployed to an edge device running in an untrusted environment.}
    \vspace{-2em}
    \label{fig:threat}
\end{figure}
Figure \ref{fig:threat} shows our threat model where the training phase is trusted but the trained model is then deployed to an inference engine that operates in an untrusted environment. The adversary is after the trained model parameters (e.g., weights and biases of a neural network)---input data privacy is out of scope \cite{wei2018know}.

We assume that the trained ML model is stored in a protected memory and the standard techniques are used to securely transfer it (i.e., bus snooping attacks are out of scope) \cite{mlIPprot}.
The adversary, therefore, has gray-box access to the device, i.e., it knows all the design details up to the level of each individual logic gate but does not know the trained ML model. 
We restrict the secret variables to just the parameters and not the hyperparameters such as the number of neurons, following earlier work~\cite{juuti2019prada,tramer2016stealing,dubey2019maskednet}. 
In fact, an adversary will still not be able to clone the model with just the hyperparameters if it does not possess the required compute power or training dataset. This is analogous to the scenario in cryptography where an adversary, even after knowing the implementation of a cipher, cannot break it without the correct key.

We target a hardware implementation of the neural network, not software. The design fully fits on the FPGA. Therefore, it does not involve any off-chip memory access and executes with constant-flow in constant time. These attributes make the design resilient to any type of digital (memory, timing, access-pattern ,etc.) side-channel attack. However, the physical side-channels like power and EM emanations still exist; we address the power-based side-channel leakages in our work. 
Other implementation attacks on neural networks such as the fault attacks \cite{breier2018practical,breier2020sniff} are out of scope.
\vspace{-.75em}
\section{Background and Related Work}
\vspace{-0.25em}
This section presents related work on the privacy of ML applications, the current state of side-channel defenses, preliminaries on BNNs, and our BNN hardware design.
\vspace{-0.75em}
\subsection{ML Model Extraction}
\vspace{-0.25em}
Recent developments in the field of ML point to several motivating scenarios that  demand asset confidentiality. 
Firstly, training is a computationally-intensive process and hence requires the model provider to invest money on high-performance compute resources (eg. a GPU cluster). The model provider might also need to invest money to buy a labeled dataset for training or label an unstructured dataset. Therefore, knowledge about either the parameters or hyperparameters can provide an unfair business advantage to the user of the model, which is why the ML model should be private. 
Theoretical model extraction analyzes the query-response pair obtained by repeatedly querying an unknown ML model to steal the parameters \cite{jagielski2019high,oh2019towards,RST19,carlini2020cryptanalytic}. 
This type of attack is similar to the class of theoretical cryptanalysis in the cryptography literature. 
Digital side-channels, by contrast, exploit the leakage of secret-data dependent \emph{intermediate computations} like access-patterns or timing in the neural network computations to steal the parameters \cite{yan2018cache,duddu2018stealing,dong2019floating,hu2019neural}, which can usually be mitigated by making the secret computations constant-flow and constant-time. Physical side-channels target the leak in the physical properties like CMOS power-draw or electromagnetic emanations that will still exist in a constant-flow/constant-time algorithm's implementation \cite{hua2018reverse,batina2018csi,wei2018know,xiang2020open,dubey2019maskednet}. Mitigating physical side-channels are thus harder than digital side-channels in hardware accelerator design and has been extensively studied in the cryptography community.

\vspace{-0.5em}
\subsection{Side-Channel Defenses}
\vspace{-0.4em}
The researchers have proposed numerous countermeasures against DPA. These countermeasures can be broadly classified as either \emph{hiding-based} or \emph{masking-based}. The former aims to make the power-consumption constant throughout the computation by using power-balancing techniques \cite{tiri2004logic,nassar2010bcdl,yu2007secure}. The latter splits the sensitive variable into multiple statistically independent shares to ensure that the power consumption is independent of the sensitive variable throughout the computation \cite{CHES:AkkGir01,CHES:GolTym02,BGK05,OMPR05,TKL05,C:IshSahWag03}. 
The security provided by hiding-based schemes hinges upon the precision of the back-end design tools to create a near-perfect power-equalized circuit by balancing the load capacitances across the leakage prone paths. This is not a trivial task and prior literature shows how a well-balanced dual-rail based defense is still vulnerable to localized EM attacks \cite{immler2017your}. By contrast, masking transforms the algorithm itself to work in a secure way by never evaluating the secret variables directly, keeping the security mostly independent of back-end design and making it a favorable choice over hiding.

\vspace{-.5em}
\subsection{Neural Network Classifiers}
\vspace{-0.4em}
Neural network algorithms learn how to perform a certain task. In the learning phase, the user sends a set of inputs and expected outputs to the machine (a.k.a., training), which helps it to approximate (or learn) the function mapping the input-output pairs. The learned function can then be used by the machine to generate outputs for unknown inputs (a.k.a., inference).

\begin{figure}
    \includegraphics[width=0.5\textwidth]{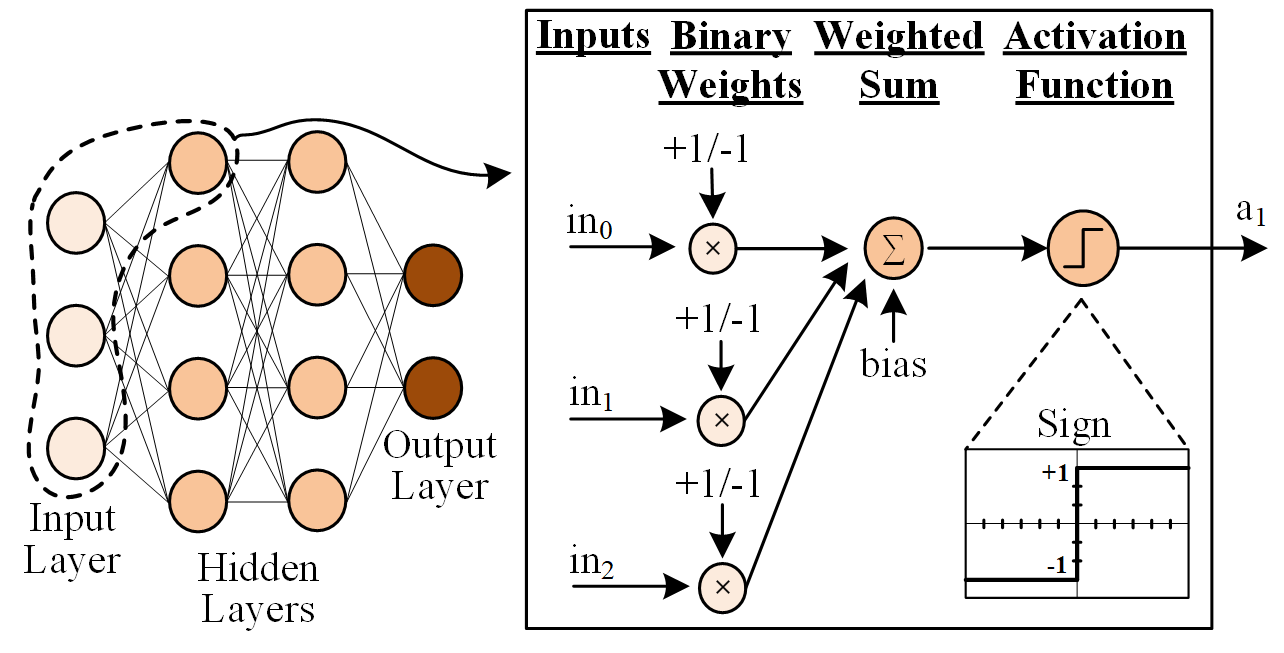}
    \vspace{-2.75em}
    \caption{A typical Binarized Neural Network where the neuron performs weighted summations on binarized weights, and the activation function is a sign function.}
    \vspace{-1.75em}
    \label{fig:neuron}
\end{figure}
A neural network consists of units called neurons (or nodes) and these neurons are usually grouped into layers. The neurons in each layer can be connected to the neurons in the previous and next layers. Each connection has a weight associated with it, which is computed as part of the training process. The neurons in a neural network work in a feed-forward fashion passing information from one layer to the next. 

The weights and biases can be initialized to be random values or a carefully chosen set before training \cite{tlearning}. 
These weights and biases are the \emph{critical parameters} that our countermeasure aims to protect.
During training, a set of inputs along with the corresponding labels are fed to the network. The network computes the error between the actual outputs and the labels and tunes the weights and biases to reduce it, converging to a state where the accuracy is acceptable. 

\vspace{-0.75em}
\subsection{Binarized Neural Networks}
\vspace{-0.25em}
The weights and biases of a neural network are typically floating-point numbers. However, high area, storage costs, and power demands of floating-point hardware do not fare well with the requirements of the resource-constrained edge devices. Fortunately, Binarized Neural Networks (BNNs) \cite{courbariaux2016binarized}, with their low hardware cost and power needs fit very well in this use-case while providing a reasonable accuracy. BNNs restrict the weights and activation to binary values (+1 and -1), which can easily be represented in hardware by a single bit. This significantly reduces the storage costs for the weights from floating-point values to binary values. The XNOR-POPCOUNT operation implemented using XNOR gates replaces the large floating-point multipliers resulting in a huge area and performance gain \cite{rastegari2016xnor}.

Figure \ref{fig:neuron} depicts the neuron computation in a fully-connected BNN. The neuron in the first hidden layer multiplies the input values with their respective binarized weights. The generated products are added to the bias, and the result is fed to the activation function, which is a sign function that binarizes the non-negative and negative inputs to +1 to -1, repectively. Hence, the activations in the subsequent layer are also binarized. 

\vspace{-.75em}
\subsection{Our Baseline BNN Hardware Design}
\vspace{-.25em}
We consider a BNN having an input layer of 784 nodes, 3 hidden layers of 1010 nodes each, and an output layer of 10 nodes. The 784 input nodes denote the 784 pixel values in the 28$\times$28 grayscale images of the Modified National Institute of Standards and Technology (MNIST) database and 10 output nodes represent the 10 output classes of the handwritten numerical digit. \cite{courbariaux2016binarized,umuroglu2017finn,rastegari2016xnor}.
\begin{figure}
    \vspace{-1em}
    \includegraphics[width=0.5\textwidth]{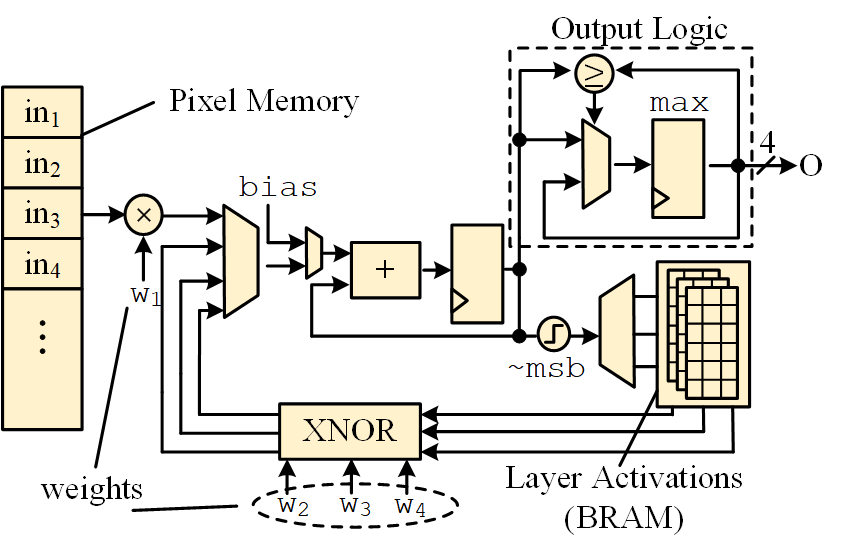}
    \vspace{-3em}
    \caption{A sequentialized hardware design of the baseline BNN using a single adder.}
    \vspace{-1.5em}
    \label{fig:base-arch}
\end{figure}

\subsubsection{Weighted Summations}
We choose to use a single adder in the design and sequentialize all the additions in the algorithm to reduce the area costs. Figure \ref{fig:base-arch} shows our baseline BNN design. The computation starts from the input layer pixel values stored in the Pixel Memory. For each node of the first hidden layer, the hardware multiplies 784 input pixel values one by one and accumulates the sum of these products. The final summation is added with the bias reusing the adder with a multiplexed input and fed to the activation function. The hardware uses XNOR and POPCOUNT\footnote{The POPCOUNT operation also involves an additional step of subtracting the number of nodes (1010) from the final sum, which can be done as part of bias addition step.} operations to perform weighted summations in the hidden layers. The final layer summations are sent to the output logic. 

\begin{figure}
  \centering
    \includegraphics[width=0.35\textwidth]{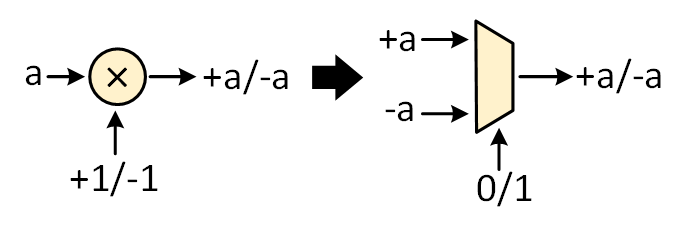}
    \vspace{-2em}
    \caption{Multiplier expressed as a multiplexer in BNNs.}
    \vspace{-1.5em}
    \label{fig:mmul}
\end{figure}

In the input layer computations, the hardware multiplies an 8-bit unsigned input pixel value with its corresponding weight. The weight values are binarized to either 0 or 1 (representing a -1 or +1, respectively). Figure \ref{fig:mmul} shows the realization of this multiplication with a multiplexer that takes in the pixel value ($a$) and its 2's complement ($-a$) as the data inputs and weight ($\pm$1) as the select line. The 8-bit unsigned pixel value, when multiplied by $\pm$1, needs to be sign-extended to 9-bits, resulting in a 9-bit multiplexer. 

\subsubsection{Activation Function}
\label{sec:actfn}
The activation function binarizes the non-negative and negative to +1 and -1 respectively for each node of the hidden layer. In hardware, this is implemented using a simple NOT gate that takes the MSB of the summations as its input.

\subsubsection{Output Layer}
The summations in the output layer represent the confidence score of each output class for the provided image. Therefore, the final classification result is the class having the maximum confidence score. Figure \ref{fig:base-arch} shows the hardware for computing the classification result. As the adder generates output layer summations, they are sent to the output logic block that performs a rolling update of the max register ($max$) if the newly received sum is greater than the previously computed max. In parallel, the hardware also stores the index of the current max node. The index stored after the final update is sent out as the final output of the neural network. The hardware takes 2.8M cycles to finish one inference.
\vspace{-0.5em}
\section{Fully Masking the Neural Network}
This section discusses the hardware design and implementation of all components in the masked neural network. 
Prior work on masking of neural networks shows that arithmetic masking alone cannot mask integer addition due to a leakage in the sign-bit \cite{dubey2019maskednet}. Hence, we apply gate-level \emph{Boolean} masking to perform integer addition in a secure fashion. We express the entire computation of the neural network as a sequence of AND and XOR operations and apply gate-level masking on the resulting expression. XORs, being linear, do not require any additional masking, and AND gates are replaced with secure, Trichina style AND gates \cite{TKL05}. Furthermore, we design specialized circuits for BNN's unique components like Masked Multiplexer and Masked Output Layer.

\begin{figure}
  \centering
    \vspace{-1em}
    \includegraphics[width=0.5\textwidth]{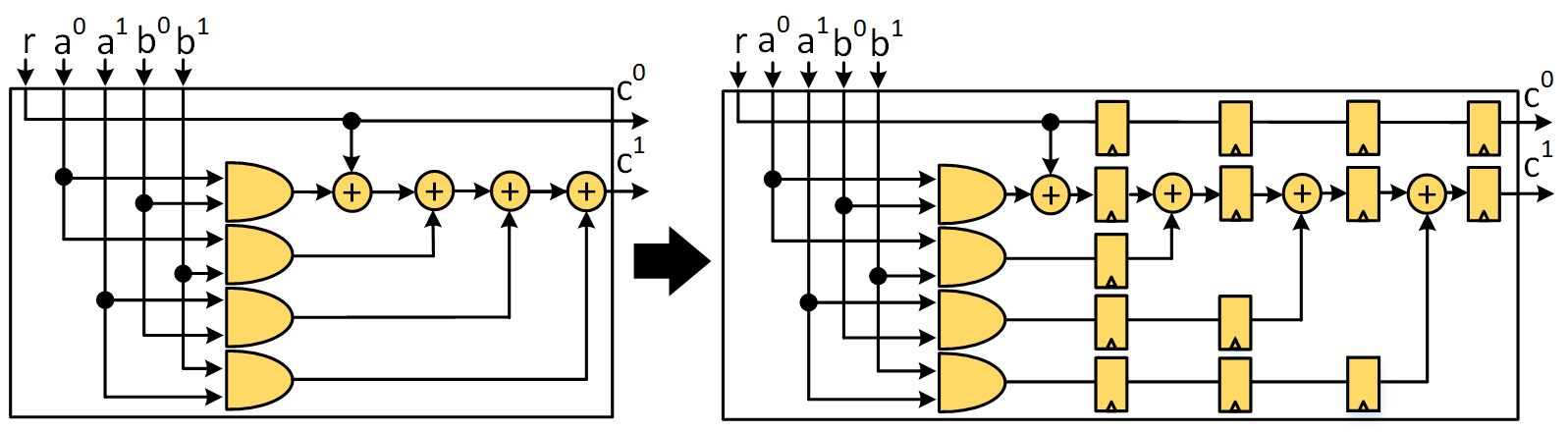}
    \vspace{-2.5em}
      \caption{Trichina's AND Gate implementation: glitch-prone (left) and glitch-resistant (right). Flip-flops synchronize arrival of signals at XOR gates' inputs to mitigate glitches.}
      \vspace{-3em}
      \label{fig:trichina}
\end{figure}

\vspace{-0.5em}
\subsection{Notations}
We first explain the notations in equations and figures. 
Any variable without a subscript or superscript represents an N-bit number. We use the subscript to refer to a single bit of the N-bit number. For example, $a_7$ refers to the $8^{th}$ bit of $a$. 
The superscript in masking refers to the different secret shares of a variable. 
To refer to a particular share of a particular bit of an N-bit number, we use both the subscript and the superscript. For example, $a_{4}^{1}$ refers to the second Boolean share of the $5^{th}$ bit of $a$. If a variable only has the superscript (say $i$), we are referring to its full N-bit $i^{th}$ Boolean share; N can also be equal to 1, in which case $a$ is simply a bit. r (or r${_i}$) denotes a fresh, random bit.
\vspace{-0.5em}
\subsection{Why Trichina's Masking Style?}
\label{ss:tgate}
Among the closely related masking styles~\cite{reparaz16-2}, we chose to implement Trichina's method due to its simplicity and implementation efficiency. 
Figure \ref{fig:trichina} (left) shows the basic structure and functionality of the Trichina's gate, which implements a 2-bit, masked, AND operation of $c=a \cdot b$.
Each input ($a$ and $b$) is split into two shares ($a^0$ and $a^1$ s.t. $a=a^0\oplus a^1$, and $b^0$ and $b^1$ s.t. $b=b^0\oplus b^1$). These shares are sequentially processed with a chain of AND gates initiated with a fresh random bit ($r$).
A single AND operation thus uses 3 random bits.
The technique ensures that output is the Boolean masked output of the original AND function, i.e., $c=c^0 \oplus c^1$, while all the intermediate computations are randomized.

Unfortunately, the straightforward adoption of Trichina's AND gate can lead to information leakage due to glitches~\cite{ICICS:NikRecRij06}. For instance, in Figure \ref{fig:trichina} (left) if the products $a_0\cdot b_0$ and $a_0\cdot b_1$ reach the input of second XOR gate before random mask $r$ reaches the input of first XOR gate, the output at the XOR gate will evaluate (glitch) to $(a_0\cdot b_0)\oplus (a_0\cdot b_1)=a_0\cdot(b_0\oplus b_1)$ temporarily, which leads to secret value $b$ being unmasked. 
Therefore, we opted for an extension of the Trichina's AND gate by adding flip-flops to synchronise the arrival of inputs at the XOR gates (see Figure \ref{fig:trichina} right). The only XOR gate not having a flip-flop at its input is the leftmost XOR gate in the path of $c_1$, which is not a problem because a glitching output at this gate does not combine two shares of the same variable. 
Similar techniques have been used in past \cite{regGlitch}. Masking styles like the Threshold gates~\cite{CHES:ManPraOsw05, maskedcmosleak,SAC:TirSch06} may be considered for even stronger security guarantees, but they will add further area-performance-randomness overhead.

\begin{figure}
  \centering
  \vspace{-1em}
    \includegraphics[width=0.5\textwidth]{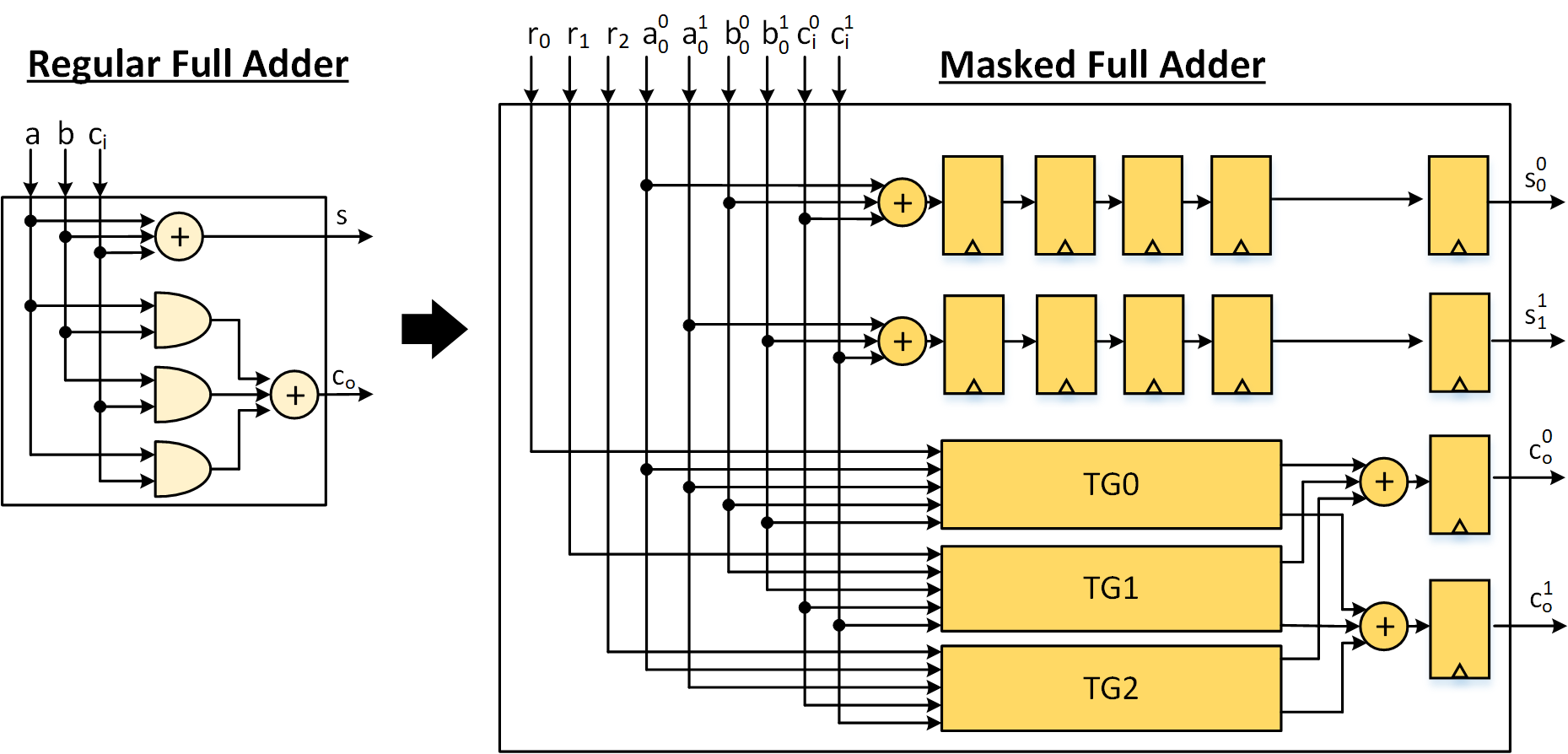}
    \vspace{-2.5em}
      \caption{Regular operation of a Full Adder (left) and its gate-level masking using Trichina AND Gates (right).}
      \vspace{-1.5em}
    \label{fig:bomanet-adders}
\end{figure}

\subsection{Masked Adder}
\vspace{-.25em}
We adopt the ripple-carry style of implementation for the adder. It is formed using N 1-bit full adders where the carry-out from each adder is the carry-in for the next adder in the chain, starting from LSB. Therefore, ripple-carry configuration eases parameterization and modular design of the Boolean masked adders.

\subsubsection{Design of a Masked Full Adder}
A 1-bit full adder takes as input two operands and a carry-in and outputs the sum and the carry, which are a function of the two operands and the carry-in. If the input operand bits are denoted by $a$ and $b$ and carry-in bit by $c$, then the Boolean equation of the sum $S$ and the carry $C$ can be described as follows:
\begin{equation}\label{eq:sum}
    S = a\oplus b\oplus c
\end{equation}
\begin{equation}\label{eq:carry}
    C = a\cdot b\oplus b\cdot c\oplus c\cdot a
\end{equation}

Figure \ref{fig:bomanet-adders}  shows the regular, 1-bit full adder (on the left), and the resulting masked adder with Trichina's AND gates (on the right). In the rest of the subsection, we will discuss the derivation of the masked full adder equations.

First step is to split the secret variables ($a$, $b$ and $c$) into Boolean shares. 
The hardware samples a fresh, random mask from a uniform distribution and performs XOR with the original variable. If we represent the random masks as $a^{0}$, $b^{0}$ and $c^{0}$, then the masked values $a^{1}$, $b^{1}$ and $c^{1}$ can be generated as follows:
\begin{equation}\label{eq:ma}
    a^1=a\oplus a^0,\;b^1=b\oplus b^0,\;c^1=c\oplus c^0
\end{equation}
A masking scheme always works on the two shares independently without ever combining them at any point in the operation. Combining the shares at any point will reconstruct the secret and create a side-channel leak at that point.

The function of sum-generation is linear, making it easy to directly and independently compute the Boolean shares of $S$:
\begin{equation*}
    S = S^0 \oplus S^1
\end{equation*}

\vspace{-1em}
\noindent{where,}
\begin{equation*}
    S^0=a^0\oplus b^0 \oplus c^0,\;S^1=a^1\oplus b^1 \oplus c^1
\end{equation*}

Unlike the sum-generation, carry-generation is a non-linear operation due to the presence of an AND operator. Hence, the hardware cannot directly and independently compute the Boolean shares $C^0$ and $C^1$ of $C$. 
We use the Trichina's construction explained in subsection \ref{ss:tgate} to mask carry-generation.

The hardware uses three Trichina's AND gates to mask the three AND operations in equation (\ref{eq:carry}) using three random masks. This generates two Boolean shares from each Trichina AND operation. At this point, the expression is linear again, and therefore, the hardware can redistribute the terms, similar to the masking of sum operation.
\begin{figure}
  \centering
    \includegraphics[width=0.5\textwidth]{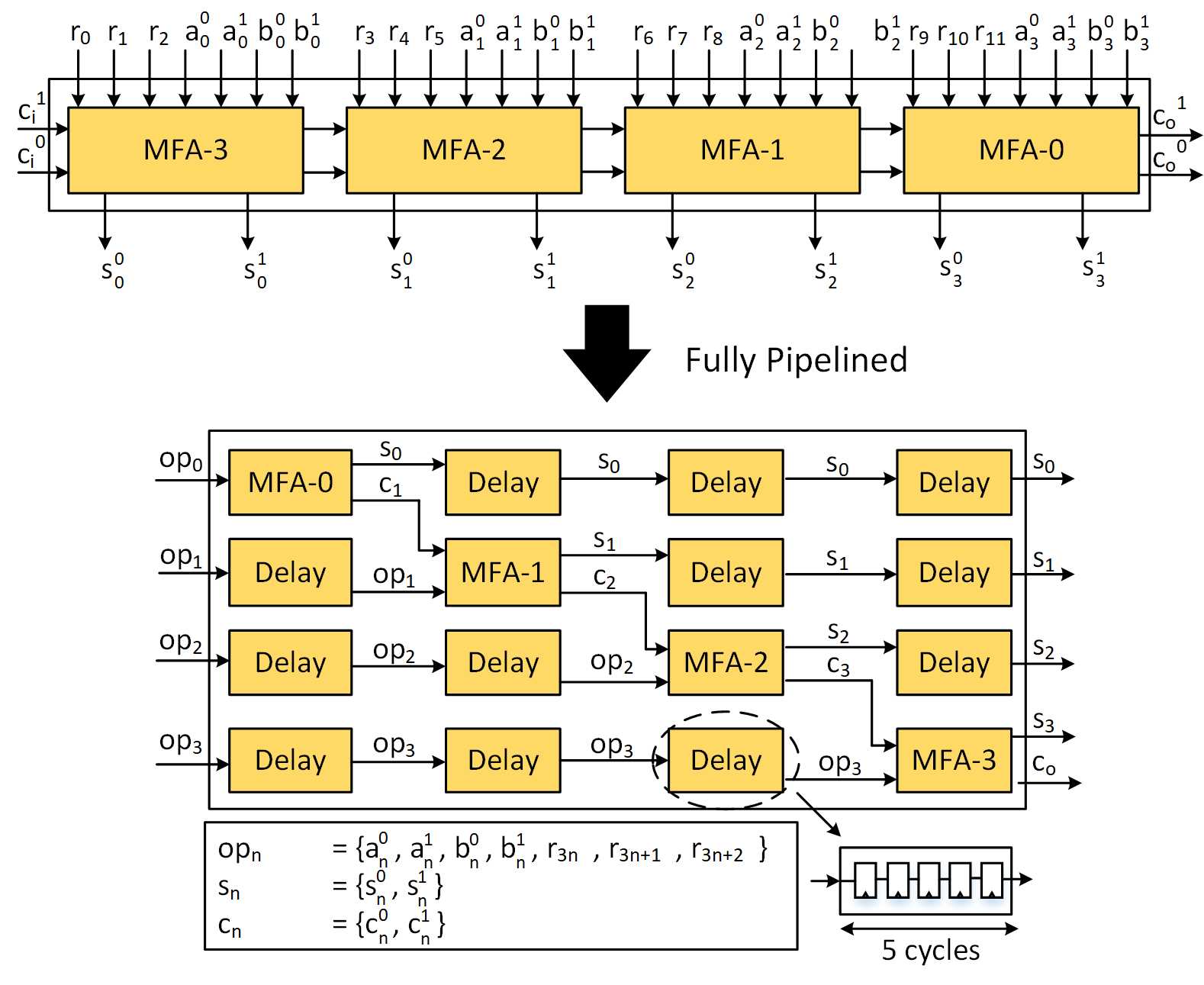}
    \vspace{-2.5em}
      \caption{Modular design of a masked 4-bit adder using Masked Full Adders (top) and its pipelined version (bottom).}
      \vspace{-1.5em}
      \label{fig:modularAdd}
\end{figure}
In the following equations, we use $TG(x,y,r)$ to represent the product $x\cdot y$ implemented via Trichina's AND Gate as illustrated in the following equation:
\begin{equation*}
    x\cdot y=TG(x,y,r) = m^0\oplus m^1
\end{equation*}
where $m^0$ and $m^1$ are the two Boolean shares of the product.
Replacing each AND operation in equation (\ref{eq:carry}) with TG, we can write
\begin{equation}\label{eq:tg0}
    TG(a,b,r_0) = d^0\oplus d^1
\end{equation}
\begin{equation}\label{eq:tg1}
    TG(b,c,r_1) = e^0\oplus e^1
\end{equation}
\begin{equation}\label{eq:tg2}
    TG(c,a,r_2) = f^0\oplus f^1
\end{equation}
where $d^0$, $d^1$, $e^0$, $e^1$, $f^0$, and $f^1$ are the output shares from each Trichina Gate. From equations (\ref{eq:carry}), (\ref{eq:tg0}), (\ref{eq:tg1}), and (\ref{eq:tg2}) we get
\begin{equation*}
    carryout = TG(a,b,r_0) \oplus TG(b,c,r_1) \oplus TG(c,a,r_2)
\end{equation*}
Replacing the TGs from equation (\ref{eq:tg0}), (\ref{eq:tg1}), and (\ref{eq:tg2}) and rearranging the terms, we get
\begin{equation*}
    carryout = (d^0\oplus e^0 \oplus f^0) \oplus (d^1 \oplus e^1 \oplus f^1)
\end{equation*}
which can also be written as a combination of two Boolean shares $C^0$ and $C^1$
\begin{equation*}
    carryout = C^0\oplus C^1
\end{equation*}

\vspace{-.5em}
\noindent{where}

\begin{equation*}
    C^0 = d^0\oplus e^0 \oplus f^0,\;C^1 = d^1 \oplus e^1 \oplus f^1
\end{equation*}
Therefore, we create a masked full adder that takes in the Boolean shares of the two bits to be added along with a carry-in and gives out the Boolean shares of the sum and carry-out. 

\subsubsection{The Modular Design of Pipelined N-bit Full Adder}
The masked full adders can be chained together to create an N-bit masked adder that can add two masked N-bit numbers. Figure \ref{fig:modularAdd} (top) shows how to construct a 4-bit masked adder as an example. We pipeline the N-bit adder to yield a throughput of one by adding registers between the full-adders corresponding to each bit (see Figure ~\ref{fig:modularAdd} (bottom)).
\vspace{-0.5em}
\subsection{Masking of Activation Function}
\begin{figure}
  \centering
    \includegraphics[width=0.45\textwidth]{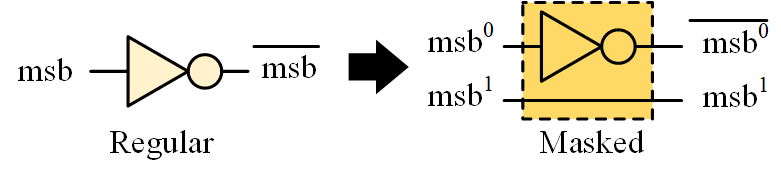}
    \vspace{-1.5em}
      \caption{While the unmasked activation function (left) is a single NOT gate, masked implementation (right) receives two Boolean shares of MSB from masked adder and inverts one of them.}
      \vspace{-1em}
      \label{fig:m-actfn}
\end{figure}

The baseline hardware implements the activation function as an inverter as discussed in \ref{sec:actfn}. In the masked version, the MSB output from the adder is a pair of Boolean shares. To perform NOT operation in a masked way, the hardware simply inverts one of the Boolean shares as Figure \ref{fig:m-actfn} shows.
\vspace{-0.5em}
\subsection{Masked Multiplexer}
\begin{figure}
  \centering
    \includegraphics[width=0.45\textwidth]{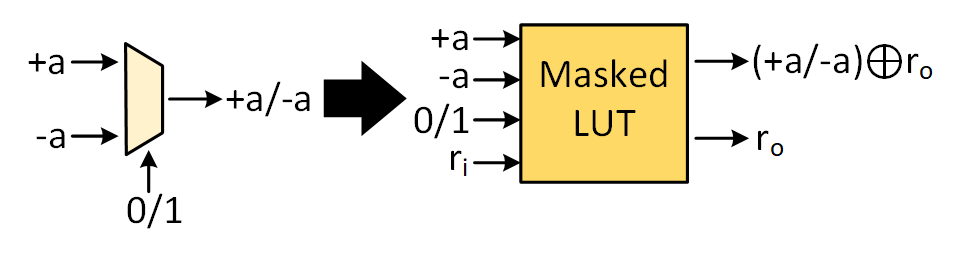}
    \vspace{-2em}
    \caption{Masking a regular multiplexer using a masked LUT taking a fresh random mask $r_i$.}
    \vspace{-1em}
    \label{fig:mux2lut}
\end{figure}
A 9-bit multiplexer is internally a set of parallel nine 1-bit multiplexers. We implement the masked 1-bit multiplexer using a 4-input 2-output masked look-up table (LUT). Figure \ref{fig:mux2lut} shows the masked LUT that takes in the original inputs ($a,-a$) and an additional fresh random mask ($r_i$) as inputs and outputs the random mask ($r_o$) which is simply the bypassed $r_i$ and the correct output XORed with the random mask. We assume that each LUT operation is atomic. Since the output functions are 4-input, 2-output, they can be mapped onto the same LUT of the target FPGA \cite{ugSpartan6}. Lesser number of inputs also obviate the need for precautions like building a carefully balanced tree of smaller input LUTs \cite{CHES:RRVV15}. Advanced masking constructions can be used to implement this function for a stronger security guarantee. As suggested in another work \cite{CHES:RRVV15}, masked look-ups can also be implemented using ROMs if the target is an ASIC, since ROMs are immutable and small in size. Thus, the (Boolean) masked output from the LUTs ensures that the secret intermediate-variable (multiplexed input pixel) always remains masked.

\vspace{-0.5em}
\subsection{Masking the Output Layer}
The hardware stores the 10 output layer summations in the form of Boolean shares. To determine the classification result, it needs to find the maximum value among the 10 masked output nodes. Specifically, it needs to \emph{compare} two signed values expressed as Boolean shares. We transform the problem of \emph{masked comparison} to \emph{masked subtraction}. 

\begin{figure}
  \centering
    \vspace{-.5em}
    \includegraphics[width=0.45\textwidth]{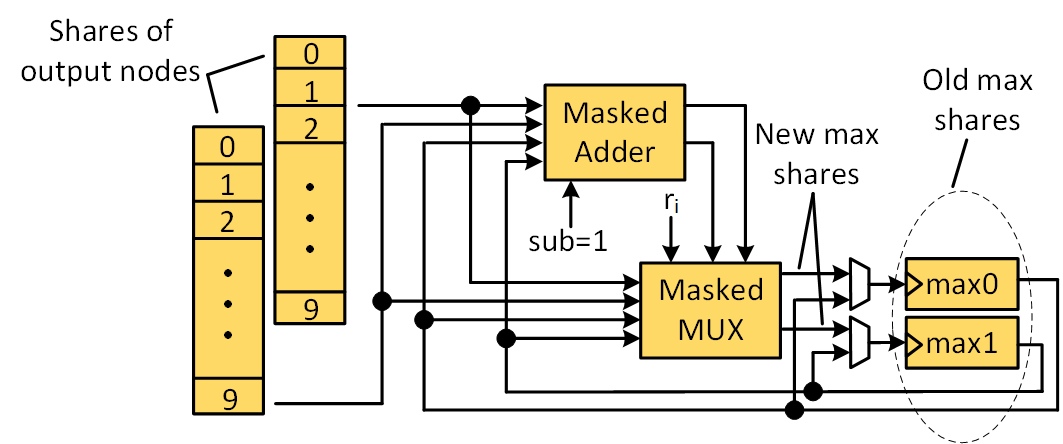}
    \vspace{-1.25em}
      \caption{Masking of the Output Layer that uses a masked subtractor and a masked multiplexer to find the node with the maximum confidence score among the 10 output nodes.}
      \label{fig:outfn}
      \vspace{-1.75em}
\end{figure}

Figure \ref{fig:outfn} shows the hardware design of the masked output layer. The hardware subtracts each output node value from the current maximum and swaps the current maximum (old max shares) with the node value (new max shares) if the MSB is 1 using a masked multiplexer. An MSB of 1 signifies that the difference is negative and hence the new sum is greater than the latest max. Instead of building a new masked subtractor, we reuse the existing masked adder to also function as a subtractor through a $sub$ flag, which is set while computing max. In parallel, the hardware uses one more masked multiplexer-based update-circuit to update the Boolean shares of the index corresponding to the current max node (not shown in the Figure). This is to prevent known-ciphertext attacks, ciphertext being the classification result in our case. Finally, the Masked Output Logic computes the classification result in the form of (Boolean) shares of the node's index having the maximum confidence score.

Subtraction is essentially adding a number with the 2's complement of another number. 2's complement is computed by taking bitwise 1's complement and adding 1 to it. A bitwise 1's complement is implemented as an XOR operation with 1 and the addition of 1 is implemented by setting the initial carry-in to be equal to 1. Since this only requires additional XOR gates, which is a linear operator, nothing changes with respect to the masking of the new adder-subtractor circuit.

\vspace{-0.5em}
\subsection{Scheduling of Operations}
\label{sec:sched}
We optimize the scheduling in such a way that the hardware maintains a throughput of 1 addition per cycle. 
The latency of the masked 20-bit adder is 100 cycles. Therefore, the result from the adder will only be available after 101 cycles (need an additional cycle for the accumulator register as well) from the time it samples the inputs. The hardware cannot feed the next input in the sequence until the previous sum is available because of the data dependency between the accumulated sum and the next accumulated sum. This incurs a stall for 101 cycles leading to a total of $784*101=79184$ cycles for each node computation. That is a $784\times$ performance drop over the unmasked implementation with a regular adder.

\begin{figure}
  \centering
    \vspace{-.5em}
    \includegraphics[width=0.5\textwidth]{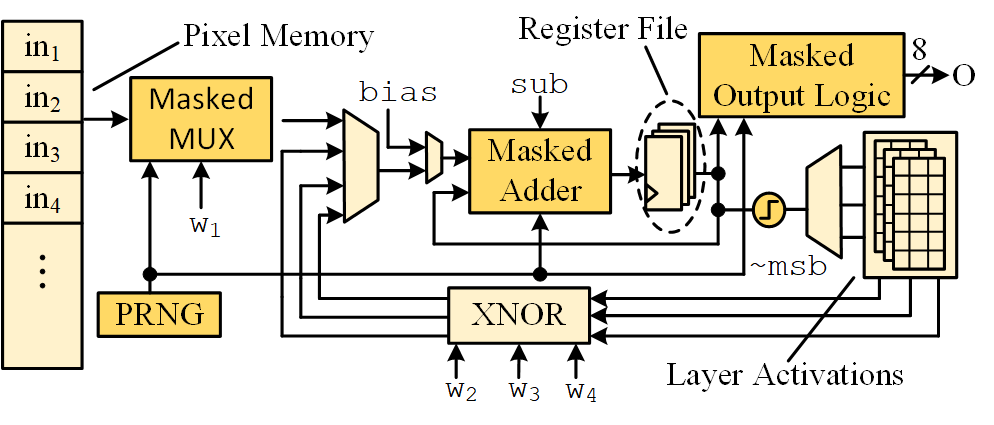}
    \vspace{-2.6em}
      \caption{Hardware Design of the Fully Masked Neural Network. The components related to masking are shown in dark yellow. The register file helps in throughput optimization by storing 101 summations parallelly.}
      \vspace{-1.65em}
      \label{fig:maskarch}
\end{figure}

 We solve the problem by finding useful work for the adder that is independent of the summation in-flight, during the stalls. We observe that computing the weighted summation of one node is completely independent of the next node's computation. The hardware utilizes this independence to improve the throughput by starting the next node computation while the result for the first node arrives. Similarly, all the nodes up till 101 can be computed upon concurrently using the same adder and achieve the exact same throughput as the baseline design. This comes at the expense of additional registers (see Figure \ref{fig:maskarch}\footnote{The register file also has a demultiplexing and multiplexing logic to update and consume the correct accumulated sum in sequence, which is not shown for simplicity.}) for storing 101 summations\footnote{This is why we use 1010 neurons, which is a multiple of 101, in the hidden layers.} plus some control logic but a throughput gain of 784$\times$ (or 1010$\times$ in hidden layers) is worthwhile. The optimization only works if the number of next-layer nodes is greater than, and a multiple of 101. This restricts optimizing the output layer (of 10 nodes) and contributes to the 3.5\% increase in the latency of the masked design.
\vspace{-0.3em}
\section{Results} 
\vspace{-.1em}
In this section, we describe the hardware setup used to implement the neural network and capture power measurements, the leakage assessment methodology that we follow to evaluate the security of the proposed design, and the hardware implementation results.

\vspace{-0.4em}
\subsection{Hardware Setup}
\vspace{-.2em}
We implement the neural network in Verilog and use Xilinx ISE 14.7 for synthesis and bitstream generation. We use the DONT\_TOUCH attribute in the code and disable the options like LUT combining, register reordering, etc. in the tool to prevent any type of optimization in the masked components. 

Our side-channel evaluation platform is the SAKURA-G FPGA board \cite{sakurag}. It hosts Xilinx Spartan-6 (XC6SLX75-2CSG484C) as the main FPGA that executes the neural network inference. An on-board amplifier amplifies the voltage drop across a 1$\Omega$ shunt resistor on the power supply line. We use Picoscope 3206D \cite{picoscope} as the oscilloscope to capture the measurements from the dedicated SMA output port of the board. The design is clocked at 24MHz and the sampling frequency of the oscilloscope is 125MHz. A higher sampling frequency leads to the challenges that we discuss in Section \ref{measDifficulty}. However, to ensure a sound evaluation, we perform first and second-order t-tests on a smaller unit of the design at a much higher precision: we conduct the experiment at a design frequency of 1.5MHz and sampling frequency of 500MHz, which translates to 333 sample points per clock cycle.

We use Riscure's Inspector SCA \cite{inspector} software to communicate with the board and initiate a capture on the oscilloscope. By default, the Inspector software does not support SAKURA-G board communication. Hence, we develop our own custom modules in the software to automate the setup. The modules implement the FPGA communication protocol and perform the register reads and writes on the FPGA to start the neural network inference and capture the classification result.

\begin{figure}
\vspace{-1em}
  \centering
    \includegraphics[width=0.5\textwidth]{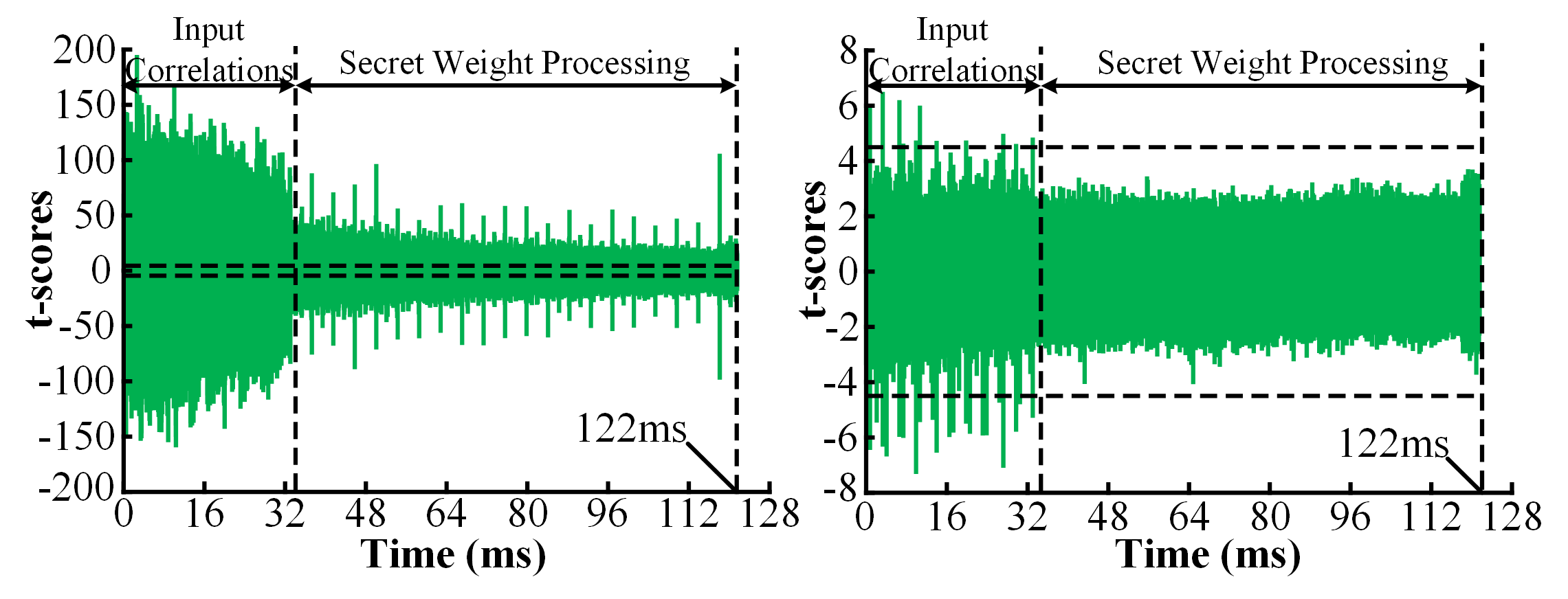}
    \vspace{-2.5em}
    \caption{TVLA results of the unmasked (left) and masked (right) implementation. The results clearly show that the unmasked design is insecure, whereas the masked design is secure with 99.99\% confidence (t-scores always below $\pm$4.5).}
    \vspace{-1em}
    \label{fig:tval_res}
\end{figure}

\begin{figure}
  \centering
    \includegraphics[width=0.45\textwidth]{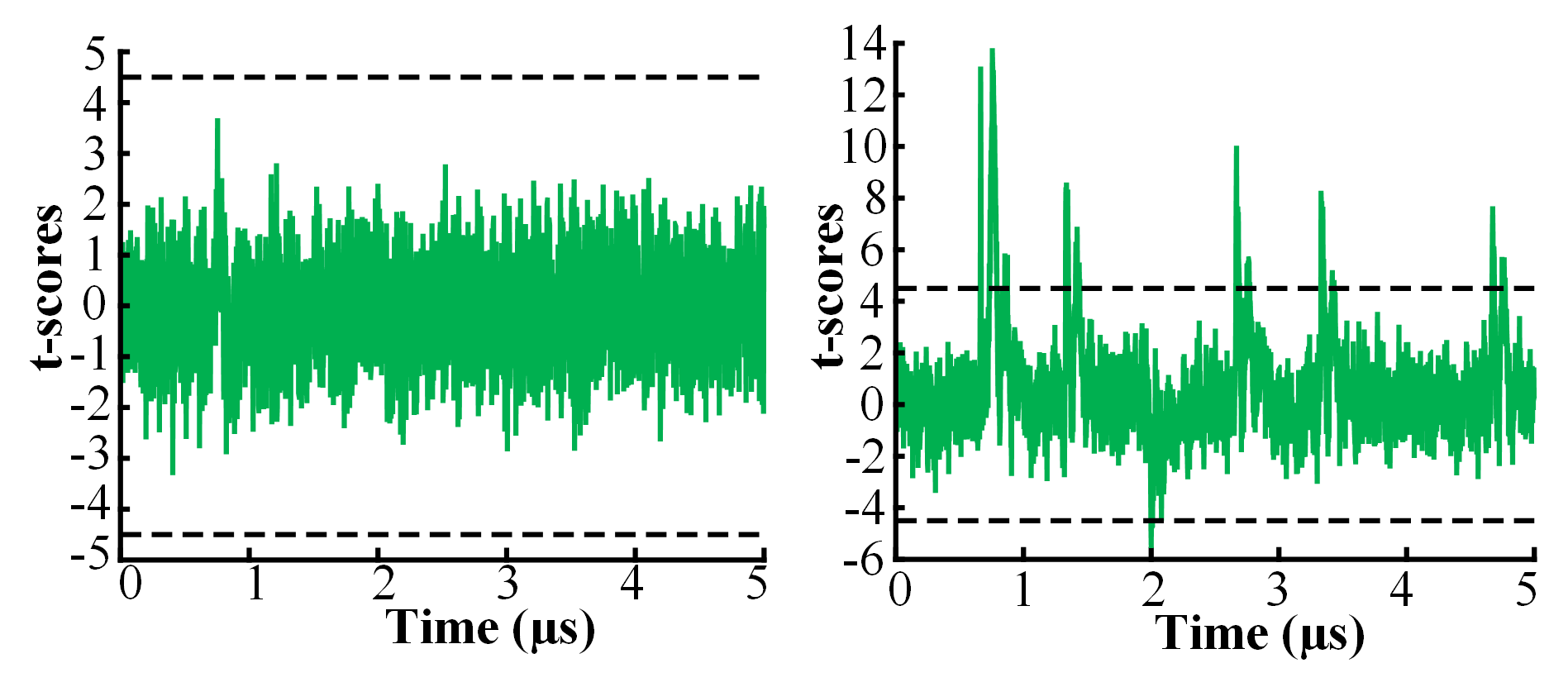}
    \vspace{-1.5em}
    \caption{First-order (left) and second-order (right) t-tests on Trichina's AND gate at a low design frequency of 1.5MHz and sampling frequency of 500MHz.}
    \vspace{-2em}
    \label{fig:tval_hp}
\end{figure}

\vspace{-.5em}
\subsection{Leakage Evaluation}
\vspace{-.25em}
We perform the leakage assessment of the proposed design using the non-specific fixed vs random t-tests, which is a common and generic way of assessing the side-channel vulnerability in a given implementation \cite{becker2013test}. 
A t-score lying within the threshold range of $\pm$4.5 implies that the power traces do not leak any information about the data being processed, with up to 99.99\% confidence. The measurement and evaluation is quite involved and we refer the reader to Section \ref{measDifficulty} for further details. We demonstrate the security up to 2M traces, which is much greater than the first-order security of the currently best-known defense that leaks at 40k traces \cite{dubey2019maskednet}. 

Pseudo Random Number Generators (PRNG) produce the fresh, random masks required for masking. 
We choose TRIVIUM~\cite{prng} as the PRNG, which is a hardware implementation friendly PRNG specified as an International Standard under ISO/IEC 29192-3, but any cryptographically-secure PRNG can be employed.
TRIVIUM generates 2$^{64}$ bits of output from an 80-bit key; hence, the PRNG has to be re-seeded before the output space is exhausted.

\vspace{-0.25em}
\subsubsection{First-order tests}
We first perform the first-order t-test on the design with PRNGs disabled, which is equivalent to an unmasked (baseline/unprotected) design. Figure \ref{fig:tval_res} (left) shows the result for this experiment where we clearly observe significant leakages since the t-scores are greater than the threshold of $\pm$4.5 for the entire execution. Then, we perform the same test, but with PRNGs switched on this time, which is equivalent to a masked design. Figure \ref{fig:tval_res} (right) shows the results for this case, where we observe that the t-scores never cross the threshold of $\pm$4.5 except the initial phase. 

The initial phase leakages are due to the input correlations during input layer computations. The hardware loads the input pixel after every 101 cycles and feeds it to the masked multiplexer. The secret variable is the \emph{weight}, which is never exposed because the masked multiplexer randomises the output using a fresh, random mask. 

\vspace{-0.25em}
\subsubsection{High Precision First and Second-order tests}
We performed univariate second-order t-test on the fully masked design \cite{schneider2016leakage}, but 1M traces were not sufficient to reveal the leakages. Due to the extremely lengthy measurement and evaluation times it was infeasible to continue the test for more number of traces. Therefore, we perform first and second-order evaluation on the isolated synchronized Trichina's AND gate, which is one of the main building blocks of the full design. We reduce the design frequency to 1.5MHz to increase the accuracy of the measurement and prevent any aliasing between clock cycles. The SNR for a single gate was not sufficient to see leakage even at 10M traces, hence we amplify the SNR by instantiating 32 independent instances of the Trichina's AND gate in the design, driven by the same inputs. We present the results for this experiment in Figure \ref{fig:tval_hp} that shows no leakage in the first-order t-test but significant leakages in the second-order t-tests for 500k traces. Thus, by ensuring success in the second-order t-tests we validate the correctness of our measurement setup and the first-order masking implementation.

\begin{table}[t!]
  \centering
  \vspace{-1em}
  \caption{Area (LUT/FF/BRAM) and Latency (in cycles) Comparison of the Unmasked and Masked Implementations.}
  \vspace{-1em}
  \begin{tabular}{ |c|c|c|c| } 
     \hline
     \textbf{Metric} & \textbf{Unmasked} & \textbf{Masked} & \textbf{Change}\\ 
     \hline
     Area & 1833/1125/163 & 9833/7624/163 & 5.3$\times$ / 6.8$\times$ / 1$\times$\\
     \hline
     Latency & \num{2.85e6} & \num{2.94e6} & 1.04$\times$\\ 
     \hline
  \end{tabular}
  \label{tab:maskOv}
  \vspace{-.8em}
\end{table}

\begin{table}[t!]
    \begin{threeparttable}
    \centering
    \caption{Block-level Area Distribution of the Unmasked and Masked Implementations (LUT/FF/BRAM)\vspace{-1.25em}}
    \begin{tabular}{ |c|c|c|c| } 
        \hline
        \textbf{Design Blocks} & \textbf{Unmasked} & \textbf{Masked} & \textbf{Fraction(\%)}\\ 
        \hline
        Adder & 10/0/0 & 954/1050/0 & 12/16/-\\ 
        \hline
        PRNGs & 0/0/0 & 1125/1314/0 & 14/20/-\\ 
        \hline
        Output Layer & 7/16/0 & 32/22/0 & 0.3/0.09/-\\ 
        \hline
        Throughput & 0/20/0 & 5337/4040/0 & 66/62/-\\
        Optimization & & &\\
        \hline
        ROMs & 411/1009/159 & 672/1009/159 & 4/-/-\\ 
        \hline
        RWMs & 0/0/4 & 0/0/4 & -\\ 
        \hline
        Misc & 486/108/0 & 1233/2605/0 & 9/38/-\\ 
        \hline
    \end{tabular}
    \label{tab:maskBlk}
    \begin{tablenotes}
        \small
        \item "-" denotes no change in the area of the unmasked and masked design.
    \end{tablenotes}
    \end{threeparttable}
    \vspace{-2em}
\end{table}

\vspace{-0.5em}
\subsection{Masking Overheads}
\vspace{-.25em}

Table~\ref{tab:maskOv} shows that the impact of masking on the performance is 1.04$\times$, and on the number of LUTs and FFs is 5.3$\times$ and 6.7$\times$ respectively. We also summarize the area contribution from each design component in Table~\ref{tab:maskBlk}. The fourth column indicates what fraction of the total increase in area (i.e., 8000 LUTs and 6499 FFs) does each component contribute. Most of the area increase is due to the throughput optimization logic---the register file accumulator logic described in subsection \ref{sec:sched}. The masked adder contributes 12\% and 16\% to the overall increase in the LUTs and FFs respectively. The increase due to the output layer logic is minimal. ROMs refer to the read-only memories storing the weights and bias values where the increase is minimal\footnote{The slight increase in the number of LUTs is because one of the memories is implemented using LUTs that might redistribute even for the same memory size.}. RWMs refer to the read-write memories storing the layer activations, which also do not show any increase as the masked version stores two bits (the Boolean shares) instead of one for the activations accommodated in the same BRAM tile. 

We compare the area-delay product (ADP) of our proposed design, BoMaNet, to MaskedNet \cite{dubey2019maskednet}, where area is defined as the sum of the number of LUTs and FFs, and delay is defined as the latency in number of cycles. The ADP of our design is \num{5e10} whereas the ADP of MaskedNET is \num{6.4e8}, which is approximately 100$\times$ lower. This is expected since MaskedNet was designed for cost-effectiveness using hiding and partial masking, but on BoMaNet every operation is masked at the gate-level to improve side-channel security. Similar overheads were observed in previous works on Boolean masking of AES~\cite{maskOverhead}.

\vspace{-.5em}
\section{Discussions}

\subsection{Proof-of-Concept vs. Optimizations}
\vspace{-.25em}
The solution we propose utilizes simple yet effective techniques to mask an inference engine. But certainly, there is scope for improvement both in terms of the hardware design and the security countermeasures. 
In this section, we discuss some possible optimizations/extensions of our work and alternate approaches taken in the field of privacy for ML.
\subsubsection{Design Optimizations}
The ripple-carry adder used in this work can be replaced with advanced adder architectures like carry-lookahead~\cite{cla}, carry-skip~\cite{csa}, or Kogge-Stone~\cite{koggestone}. These architectures commonly possess an additional logic block that pre-computes the \emph{generate} and \emph{propagate} bits. Therefore, additional randomness will be needed to mask the non-linear \emph{generate} expression. 
All these adders have more combinational logic than the ripple-carry adder, which may make it harder to avoid glitches. To that end, prior work on TI-based secure versions of ripple-carry and Kogge-Stone adders can be extended~\cite{boma-adder15}. Another potential optimization is the use of other masking styles like DoM~\cite{gross2016domain} or manual techniques \cite{manualGlitch} to reduce the area and randomness overheads.

\subsubsection{Limitations}
We reduce glitch-related vulnerabilities using registers at each stage, which is a low-cost, practical solution. Other works have proposed stronger countermeasures, at the cost of higher performance and area overheads~\cite{ICICS:NikRecRij06,gross2016domain}. 
The quest for stronger and more efficient countermeasures is never-ending; masking of AES is still being explored, even 20 years after the initial work~\cite{CHES:AkkGir01}, due to the advent of more efficient or secure masking schemes \cite{d+1shares} and more potent attacks~\cite{moos2017static,TKL05}. 

Our solution is first-order secure but there is scope for construction of higher-order masked versions. However, higher-order security is a delicate task; Moos \textit{et al.} recently showed that a straightforward extension of masking schemes to higher-order suffers from both local and compositional flaws \cite{TCHES:MMSS19} and masking extensions were proposed in another recent work \cite{cassiershardware}. 

This is the first work on fully-masked neural networks and we foresee follow ups as we have experienced in the cryptographic research of AES masking, even after 20 years of intensive study.

\vspace{-0.5em}
\subsection{Measurement Challenges}
\vspace{-0.25em}
\label{measDifficulty}
We faced some unique challenges that are not generally seen with the symmetric-key cryptographic evaluations. Inference becomes a lengthy operation, especially for an area-optimized design---the inference latency of our design is roughly 3 million cycles. For a design frequency of 24MHz, the execution time translates to 122ms per inference. If the oscilloscope samples at 125MHz (sample interval of 8ns) the number of sample points to be captured per power trace is equal to 15 million. This significantly slows down the capturing of power traces. In our case, capturing 2 million power traces took one week, which means capturing 100 million traces as AES evaluation~\cite{d+1shares} will take roughly a year. Performing TVLA on such large traces (~28TB, in our case) also takes a significant amount of time: it took 3 days to get one t-score plot during our evaluations on a high-end PC\footnote{Intel Core i9-9900K, 64GB RAM.}. One possibility to avoid this problem is looking at a small subset of representative traces of the computation~\cite{DLRSA}, but, we instead conduct a comprehensive evaluation of our design.

\vspace{-0.5em}
\subsection{Theoretical vs Side-Channel Attacks}
\vspace{-0.25em}
Theoretical model extraction by training a proxy model on a synthetically generated dataset using the predictions from the unknown victim model is an active area of research \cite{jagielski2019high,carlini2020cryptanalytic}. These attacks mostly assume a black-box access to the model and successfully extract the model parameters after a certain number of queries. This number ranges typically in the order of $2^{20}$ \cite{carlini2020cryptanalytic}. By contrast, physical side-channel attacks only require a few thousand queries to successfully steal all the parameters \cite{dubey2019maskednet}. This is partly due to fact that physical side-channel attacks can extract information about intermediate computations even in a black-box setting. Physical side-channel attacks also do not require the generation of the synthetic dataset, unlike most theoretical attacks.

\vspace{-.5em}
\subsection{Orthogonal ML Defences}
\vspace{-.25em}
There has been some work on defending the ML models against stealing of inputs and parameters using other techniques like Homomorphic Encryption (HE) and Secure Multi-Party Computation (SMPC) \cite{gazelle18,xonn19,delphi20}, Watermarking \cite{rouhani2018deepsigns,adi2018turning}, and Trusted Execution Engines (TEE) \cite{preventingnn18,tramer2018slalom,mlcapsule18}. The survey by Isakov \textit{et al.} and the draft by NIST is a good reference for a more exhaustive list \cite{mlNIST,isakov2019survey}. The computational needs of HE might not be suitable for edge computing. The current SMPC defenses predominantly target a cloud-based ML framework, not edge. We propose masking, which is an extension of SMPC on hardware and we believe that it is a promising direction for ML side-channel defenses as it has been on cryptographic applications. Watermarking techniques are punitive methods that cannot prevent physical side-channel attacks. TEEs are subject to ever-evolving microarchitectural attacks and typically are not available in edge/IoT nodes.

\section{Conclusions and Future Outlook}
Physical side-channel analysis of neural networks is a new, promising direction in hardware security where the attacks are rapidly evolving compared to defenses.
This work proposed the first fully-masked neural network, demonstrated the security with up to 2M traces, and quantified the overheads of a potential countermeasure.
We have addressed the key challenge of masking arithmetic shares of integer addition~\cite{dubey2019maskednet} through Boolean masking. 
We furthermore presented ideas on how to mask the unique linear and non-linear computations of a fully-connected neural network that do not exist in cryptographic applications.

The large variety in neural network architectures in terms of the level of quantization, the types of layer operations (e.g., Convolution, Maxpool, Softmax), and the types of activation functions (e.g., ReLU, Sigmoid, Tanh) presents a large design space for neural network side-channel defenses. 
This paper focused on BNNs as they are a good starting point.
The ideas presented in this work serve as a benchmark to analyze the vulnerabilities that exist in neural network computations and to construct more robust and efficient countermeasures.

\section{Acknowledgements}
This project is supported in part by NSF under Grants No. 1943245 and SRC GRC Task 2908.001. We also thank Dr Vikram Suresh and Dr Sohrab Aftabjahani for their valuable feedback on the design.

\newpage
\vspace{-2em}
\bibliographystyle{ACM-Reference-Format}
\bibliography{papers.bib}


\begin{thebibliography}{78}


\ifx \showCODEN    \undefined \def \showCODEN     #1{\unskip}     \fi
\ifx \showDOI      \undefined \def \showDOI       #1{#1}\fi
\ifx \showISBNx    \undefined \def \showISBNx     #1{\unskip}     \fi
\ifx \showISBNxiii \undefined \def \showISBNxiii  #1{\unskip}     \fi
\ifx \showISSN     \undefined \def \showISSN      #1{\unskip}     \fi
\ifx \showLCCN     \undefined \def \showLCCN      #1{\unskip}     \fi
\ifx \shownote     \undefined \def \shownote      #1{#1}          \fi
\ifx \showarticletitle \undefined \def \showarticletitle #1{#1}   \fi
\ifx \showURL      \undefined \def \showURL       {\relax}        \fi
\providecommand\bibfield[2]{#2}
\providecommand\bibinfo[2]{#2}
\providecommand\natexlab[1]{#1}
\providecommand\showeprint[2][]{arXiv:#2}

\bibitem[\protect\citeauthoryear{Adi~et al.}{Adi~et al.}{2018}]%
        {adi2018turning}
\bibfield{author}{\bibinfo{person}{Yossi Adi~et al.}}
  \bibinfo{year}{2018}\natexlab{}.
\newblock \showarticletitle{Turning Your Weakness Into a Strength: Watermarking
  Deep Neural Networks by Backdooring}. In \bibinfo{booktitle}{\emph{USENIX
  Security '18}}.
\newblock


\bibitem[\protect\citeauthoryear{Becker~et al.}{Becker~et al.}{2013}]%
        {becker2013test}
\bibfield{author}{\bibinfo{person}{George Becker~et al.}}
  \bibinfo{year}{2013}\natexlab{}.
\newblock \showarticletitle{Test Vector Leakage Assessment (TVLA) methodology
  in practice}. In \bibinfo{booktitle}{\emph{International Cryptographic Module
  Conference}}, Vol.~\bibinfo{volume}{1001}.
\newblock


\bibitem[\protect\citeauthoryear{Bl{\"o}mer~et al.}{Bl{\"o}mer~et al.}{2005}]%
        {BGK05}
\bibfield{author}{\bibinfo{person}{Johannes Bl{\"o}mer~et al.}}
  \bibinfo{year}{2005}\natexlab{}.
\newblock \showarticletitle{Provably Secure Masking of AES}. In
  \bibinfo{booktitle}{\emph{Selected Areas in Cryptography}}.
\newblock


\bibitem[\protect\citeauthoryear{Breier~et al.}{Breier~et al.}{2018}]%
        {breier2018practical}
\bibfield{author}{\bibinfo{person}{Jakub Breier~et al.}}
  \bibinfo{year}{2018}\natexlab{}.
\newblock \showarticletitle{Practical Fault Attack on Deep Neural Networks}. In
  \bibinfo{booktitle}{\emph{2018 ACM SIGSAC Conference on Computer and
  Communications Security}}.
\newblock


\bibitem[\protect\citeauthoryear{Breier~et al.}{Breier~et al.}{2020}]%
        {breier2020sniff}
\bibfield{author}{\bibinfo{person}{Jakub Breier~et al.}}
  \bibinfo{year}{2020}\natexlab{}.
\newblock \showarticletitle{{SNIFF}: Reverse Engineering of Neural Networks
  with Fault Attacks}.
\newblock \bibinfo{journal}{\emph{arXiv preprint arXiv:2002.11021}}
  (\bibinfo{year}{2020}).
\newblock


\bibitem[\protect\citeauthoryear{Carbone~et al.}{Carbone~et al.}{2019}]%
        {DLRSA}
\bibfield{author}{\bibinfo{person}{Mathieu Carbone~et al.}}
  \bibinfo{year}{2019}\natexlab{}.
\newblock \showarticletitle{Deep Learning to Evaluate Secure RSA
  Implementations}.
\newblock \bibinfo{journal}{\emph{TCHES}}  \bibinfo{volume}{2019}
  (\bibinfo{year}{2019}).
\newblock


\bibitem[\protect\citeauthoryear{Chen~et al.}{Chen~et al.}{2015}]%
        {chen2015differential}
\bibfield{author}{\bibinfo{person}{Cong Chen~et al.}}
  \bibinfo{year}{2015}\natexlab{}.
\newblock \showarticletitle{Differential Power Analysis of a {McEliece}
  Cryptosystem}. In \bibinfo{booktitle}{\emph{International Conference on
  Applied Cryptography and Network Security}}. Springer.
\newblock


\bibitem[\protect\citeauthoryear{Courbariaux~et al.}{Courbariaux~et
  al.}{2016}]%
        {courbariaux2016binarized}
\bibfield{author}{\bibinfo{person}{Matthieu Courbariaux~et al.}}
  \bibinfo{year}{2016}\natexlab{}.
\newblock \showarticletitle{Binarized Neural Networks: Training Deep Neural
  Networks with Weights and Activations Constrained to +1 or-1}.
\newblock  (\bibinfo{year}{2016}).
\newblock


\bibitem[\protect\citeauthoryear{Das~et al.}{Das~et al.}{2019}]%
        {das2019stellar}
\bibfield{author}{\bibinfo{person}{Debayan Das~et al.}}
  \bibinfo{year}{2019}\natexlab{}.
\newblock \showarticletitle{STELLAR: A Generic EM Side-Channel Attack
  Protection through Ground-Up Root-cause Analysis}. In
  \bibinfo{booktitle}{\emph{HOST '19}}.
\newblock


\bibitem[\protect\citeauthoryear{De~Canni{\`e}re~et al.}{De~Canni{\`e}re~et
  al.}{2008}]%
        {prng}
\bibfield{author}{\bibinfo{person}{Christophe De~Canni{\`e}re~et al.}}
  \bibinfo{year}{2008}\natexlab{}.
\newblock \showarticletitle{Trivium}.
\newblock In \bibinfo{booktitle}{\emph{New Stream Cipher Designs}}.
\newblock


\bibitem[\protect\citeauthoryear{Dong~et al.}{Dong~et al.}{2019}]%
        {dong2019floating}
\bibfield{author}{\bibinfo{person}{Gaofeng Dong~et al.}}
  \bibinfo{year}{2019}\natexlab{}.
\newblock \showarticletitle{Floating-Point Multiplication Timing Attack on Deep
  Neural Network}. In \bibinfo{booktitle}{\emph{IEEE International Conference
  on Smart Internet of Things}}.
\newblock


\bibitem[\protect\citeauthoryear{Doran}{Doran}{1988}]%
        {cla}
\bibfield{author}{\bibinfo{person}{Robert~W Doran}.}
  \bibinfo{year}{1988}\natexlab{}.
\newblock \showarticletitle{Variants of an improved carry look-ahead adder}.
\newblock \bibinfo{journal}{\emph{IEEE Trans. Comput.}} \bibinfo{volume}{37},
  \bibinfo{number}{9} (\bibinfo{year}{1988}), \bibinfo{pages}{1110--1113}.
\newblock


\bibitem[\protect\citeauthoryear{Dubey~et al.}{Dubey~et al.}{2019}]%
        {dubey2019maskednet}
\bibfield{author}{\bibinfo{person}{Anuj Dubey~et al.}}
  \bibinfo{year}{2019}\natexlab{}.
\newblock \showarticletitle{{MaskedNet}: A Pathway for Secure Inference against
  Power Side-Channel Attacks}.
\newblock \bibinfo{journal}{\emph{arXiv preprint arXiv:1910.13063}}
  (\bibinfo{year}{2019}).
\newblock


\bibitem[\protect\citeauthoryear{Duddu~et al.}{Duddu~et al.}{2018}]%
        {duddu2018stealing}
\bibfield{author}{\bibinfo{person}{Vasisht Duddu~et al.}}
  \bibinfo{year}{2018}\natexlab{}.
\newblock \showarticletitle{Stealing Neural Networks via Timing Side Channels}.
\newblock \bibinfo{journal}{\emph{arXiv preprint arXiv:1812.11720}}
  (\bibinfo{year}{2018}).
\newblock


\bibitem[\protect\citeauthoryear{et~al.}{et~al.}{2012}]%
        {manualGlitch}
\bibfield{author}{\bibinfo{person}{Amir~{Moradi} et al.}}
  \bibinfo{year}{2012}\natexlab{}.
\newblock \showarticletitle{Glitch-Free Implementation of Masking in Modern
  FPGAs}. In \bibinfo{booktitle}{\emph{HOST '12}}.
\newblock


\bibitem[\protect\citeauthoryear{et~al.}{et~al.}{2018a}]%
        {TCHES:PSKH18}
\bibfield{author}{\bibinfo{person}{Aesun~Park et al.}}
  \bibinfo{year}{2018}\natexlab{a}.
\newblock \showarticletitle{Side-Channel Attacks on Post-Quantum Signature
  Schemes based on Multivariate Quadratic Equations}.
\newblock \bibinfo{journal}{\emph{TCHES}} \bibinfo{volume}{2018},
  \bibinfo{number}{3} (\bibinfo{year}{2018}).
\newblock


\bibitem[\protect\citeauthoryear{et~al.}{et~al.}{2020a}]%
        {cassiershardware}
\bibfield{author}{\bibinfo{person}{Gaëtan~Cassiers et al.}}
  \bibinfo{year}{2020}\natexlab{a}.
\newblock \showarticletitle{Hardware Private Circuits: From Trivial Composition
  to Full Verification}.
\newblock \bibinfo{howpublished}{ePrint, Report 2020/185}.
\newblock  (\bibinfo{year}{2020}).
\newblock


\bibitem[\protect\citeauthoryear{et~al.}{et~al.}{2014}]%
        {sakurag}
\bibfield{author}{\bibinfo{person}{H.~{Guntur} et al.}}
  \bibinfo{year}{2014}\natexlab{}.
\newblock \showarticletitle{Side-channel AttacK User Reference Architecture
  board SAKURA-G}. In \bibinfo{booktitle}{\emph{2014 IEEE 3rd Global Conference
  on Consumer Electronics (GCCE)}}.
\newblock


\bibitem[\protect\citeauthoryear{et~al.}{et~al.}{2007}]%
        {SAC:TirSch06}
\bibfield{author}{\bibinfo{person}{Kris~Tiri et al.}}
  \bibinfo{year}{2007}\natexlab{}.
\newblock \showarticletitle{Changing the Odds Against Masked Logic}. In
  \bibinfo{booktitle}{\emph{SAC 2006: 13th Annual International Workshop on
  Selected Areas in Cryptography}}, Vol.~\bibinfo{volume}{4356}.
\newblock


\bibitem[\protect\citeauthoryear{et~al.}{et~al.}{2019a}]%
        {batina2018csi}
\bibfield{author}{\bibinfo{person}{Lejla~Batina et al.}}
  \bibinfo{year}{2019}\natexlab{a}.
\newblock \showarticletitle{{CSI} {NN}: Reverse Engineering of Neural Network
  Architectures Through Electromagnetic Side Channel}. In
  \bibinfo{booktitle}{\emph{{USENIX} Security '19}}.
\newblock


\bibitem[\protect\citeauthoryear{et~al.}{et~al.}{2009}]%
        {regGlitch}
\bibfield{author}{\bibinfo{person}{Monjur~{Alam} et al.}}
  \bibinfo{year}{2009}\natexlab{}.
\newblock \showarticletitle{Effect of glitches against masked AES S-box
  implementation and countermeasure}.
\newblock \bibinfo{journal}{\emph{IET Information Security}}
  (\bibinfo{year}{2009}).
\newblock


\bibitem[\protect\citeauthoryear{et~al.}{et~al.}{2019b}]%
        {jagielski2019high}
\bibfield{author}{\bibinfo{person}{Matthew~Jagielski et al.}}
  \bibinfo{year}{2019}\natexlab{b}.
\newblock \bibinfo{title}{High Accuracy and High Fidelity Extraction of Neural
  Networks}.
\newblock
\newblock
\showeprint[arxiv]{cs.LG/1909.01838}


\bibitem[\protect\citeauthoryear{et~al.}{et~al.}{2001}]%
        {CHES:AkkGir01}
\bibfield{author}{\bibinfo{person}{Mehdi-Laurent~Akkar et al.}}
  \bibinfo{year}{2001}\natexlab{}.
\newblock \showarticletitle{An Implementation of {DES} and {AES}, Secure
  against Some Attacks}. In \bibinfo{booktitle}{\emph{{CHES}~2001}},
  Vol.~\bibinfo{volume}{2162}. \bibinfo{publisher}{Springer, Heidelberg,
  Germany}.
\newblock


\bibitem[\protect\citeauthoryear{et~al.}{et~al.}{2020b}]%
        {carlini2020cryptanalytic}
\bibfield{author}{\bibinfo{person}{Nicholas~Carlini et al.}}
  \bibinfo{year}{2020}\natexlab{b}.
\newblock \bibinfo{title}{Cryptanalytic Extraction of Neural Network Models}.
\newblock
\newblock
\showeprint[arxiv]{cs.LG/2003.04884}


\bibitem[\protect\citeauthoryear{et~al.}{et~al.}{2015}]%
        {CHES:RRVV15}
\bibfield{author}{\bibinfo{person}{Oscar~Reparaz et al.}}
  \bibinfo{year}{2015}\natexlab{}.
\newblock \showarticletitle{A Masked Ring-{LWE} Implementation}. In
  \bibinfo{booktitle}{\emph{{CHES}~2015}}.
\newblock


\bibitem[\protect\citeauthoryear{et~al}{et~al}{1999}]%
        {C:KocJafJun99}
\bibfield{author}{\bibinfo{person}{Paul C.~Kocher et al}.}
  \bibinfo{year}{1999}\natexlab{}.
\newblock \showarticletitle{Differential Power Analysis}. In
  \bibinfo{booktitle}{\emph{Advances in Cryptology -- {CRYPTO}'99}}.
  \bibinfo{publisher}{Springer, Heidelberg, Germany}.
\newblock


\bibitem[\protect\citeauthoryear{et~al.}{et~al.}{2018b}]%
        {mlIPprot}
\bibfield{author}{\bibinfo{person}{Rosario~{Cammarota} et al.}}
  \bibinfo{year}{2018}\natexlab{b}.
\newblock \showarticletitle{Machine Learning IP Protection}. In
  \bibinfo{booktitle}{\emph{ICCAD '18}}.
\newblock


\bibitem[\protect\citeauthoryear{et~al.}{et~al.}{2010}]%
        {tlearning}
\bibfield{author}{\bibinfo{person}{Sinno Jialin~Pan et al.}}
  \bibinfo{year}{2010}\natexlab{}.
\newblock \showarticletitle{A Survey on Transfer Learning}.
\newblock \bibinfo{journal}{\emph{IEEE Transactions on Knowledge and Data
  Engineering}} (\bibinfo{year}{2010}).
\newblock


\bibitem[\protect\citeauthoryear{et~al.}{et~al.}{2005}]%
        {CHES:ManPraOsw05}
\bibfield{author}{\bibinfo{person}{Stefan~Mangard et al.}}
  \bibinfo{year}{2005}\natexlab{}.
\newblock \showarticletitle{Successfully Attacking Masked {AES} Hardware
  Implementations}. In \bibinfo{booktitle}{\emph{{CHES}~2005}}.
\newblock


\bibitem[\protect\citeauthoryear{et~al.}{et~al.}{2006}]%
        {ICICS:NikRecRij06}
\bibfield{author}{\bibinfo{person}{Svetla~Nikova et al.}}
  \bibinfo{year}{2006}\natexlab{}.
\newblock \showarticletitle{Threshold Implementations Against Side-Channel
  Attacks and Glitches}. In \bibinfo{booktitle}{\emph{ICICS '06}}.
\newblock


\bibitem[\protect\citeauthoryear{et~al.}{et~al.}{2016}]%
        {d+1shares}
\bibfield{author}{\bibinfo{person}{Thomas De~Cnudde et al.}}
  \bibinfo{year}{2016}\natexlab{}.
\newblock \bibinfo{title}{Masking AES with d+1 Shares in Hardware}.
\newblock \bibinfo{howpublished}{IACR ePrint, 2016/631}.
\newblock


\bibitem[\protect\citeauthoryear{et~al.}{et~al.}{2019c}]%
        {TCHES:MMSS19}
\bibfield{author}{\bibinfo{person}{Thorben~Moos et al.}}
  \bibinfo{year}{2019}\natexlab{c}.
\newblock \showarticletitle{Glitch-Resistant Masking Revisited}.
\newblock \bibinfo{journal}{\emph{{IACR} TCHES}} \bibinfo{volume}{2019},
  \bibinfo{number}{2} (\bibinfo{year}{2019}).
\newblock


\bibitem[\protect\citeauthoryear{et~al.}{et~al.}{2019d}]%
        {hu2019neural}
\bibfield{author}{\bibinfo{person}{Xing~Hu et al.}}
  \bibinfo{year}{2019}\natexlab{d}.
\newblock \bibinfo{title}{Neural Network Model Extraction Attacks in Edge
  Devices by Hearing Architectural Hints}.
\newblock
\newblock
\showeprint[arxiv]{cs.CR/1903.03916}


\bibitem[\protect\citeauthoryear{et~al.}{et~al.}{2003}]%
        {C:IshSahWag03}
\bibfield{author}{\bibinfo{person}{Yuval~Ishai et al.}}
  \bibinfo{year}{2003}\natexlab{}.
\newblock \showarticletitle{Private Circuits: Securing Hardware against Probing
  Attacks}. In \bibinfo{booktitle}{\emph{Advances in Cryptology --
  {CRYPTO}~2003}}, Vol.~\bibinfo{volume}{2729}.
\newblock


\bibitem[\protect\citeauthoryear{et~al.}{et~al.}{2018c}]%
        {maskOverhead}
\bibfield{author}{\bibinfo{person}{Yuan~{Yao} et al.}}
  \bibinfo{year}{2018}\natexlab{c}.
\newblock \showarticletitle{Fault-Assisted Side-Channel Analysis of Masked
  Implementations}. In \bibinfo{booktitle}{\emph{HOST '18}}.
\newblock


\bibitem[\protect\citeauthoryear{Goli{\'c}~et al.}{Goli{\'c}~et al.}{2003}]%
        {CHES:GolTym02}
\bibfield{author}{\bibinfo{person}{Jovan~D Goli{\'c}~et al.}}
  \bibinfo{year}{2003}\natexlab{}.
\newblock \showarticletitle{Multiplicative Masking and Power Analysis of
  {AES}}. In \bibinfo{booktitle}{\emph{{CHES}~2002}},
  Vol.~\bibinfo{volume}{2523}. \bibinfo{publisher}{Springer, Heidelberg,
  Germany}.
\newblock


\bibitem[\protect\citeauthoryear{Gro{\ss}~et al.}{Gro{\ss}~et al.}{2016}]%
        {gross2016domain}
\bibfield{author}{\bibinfo{person}{Hannes Gro{\ss}~et al.}}
  \bibinfo{year}{2016}\natexlab{}.
\newblock \showarticletitle{Domain-Oriented Masking: Compact Masked Hardware
  Implementations with Arbitrary Protection Order}.
\newblock \bibinfo{journal}{\emph{IACR ePrint}} (\bibinfo{year}{2016}).
\newblock


\bibitem[\protect\citeauthoryear{Hanzlik~et al.}{Hanzlik~et al.}{2018}]%
        {mlcapsule18}
\bibfield{author}{\bibinfo{person}{Lucjan Hanzlik~et al.}}
  \bibinfo{year}{2018}\natexlab{}.
\newblock \showarticletitle{{MLCapsule}: Guarded Offline Deployment of Machine
  Learning as a Service}.
\newblock \bibinfo{journal}{\emph{arXiv preprint arXiv:1808.00590}}
  (\bibinfo{year}{2018}).
\newblock


\bibitem[\protect\citeauthoryear{Hua~et al.}{Hua~et al.}{2018}]%
        {hua2018reverse}
\bibfield{author}{\bibinfo{person}{Weizhe Hua~et al.}}
  \bibinfo{year}{2018}\natexlab{}.
\newblock \showarticletitle{Reverse Engineering Convolutional Neural Networks
  through Side-Channel Information Leaks}. In \bibinfo{booktitle}{\emph{DAC
  '18}}.
\newblock


\bibitem[\protect\citeauthoryear{Immler~et al.}{Immler~et al.}{2017}]%
        {immler2017your}
\bibfield{author}{\bibinfo{person}{Vincent Immler~et al.}}
  \bibinfo{year}{2017}\natexlab{}.
\newblock \showarticletitle{Your Rails Cannot Hide From Localized EM: How
  Dual-Rail Logic Fails on FPGAs}. In \bibinfo{booktitle}{\emph{CHES 2017}}.
\newblock


\bibitem[\protect\citeauthoryear{Isakov~et al.}{Isakov~et al.}{2019}]%
        {isakov2019survey}
\bibfield{author}{\bibinfo{person}{Mihailo Isakov~et al.}}
  \bibinfo{year}{2019}\natexlab{}.
\newblock \showarticletitle{Survey of Attacks and Defenses on Edge-Deployed
  Neural Networks}. In \bibinfo{booktitle}{\emph{HPEC '19}}.
\newblock


\bibitem[\protect\citeauthoryear{Isakovet~al.}{Isakovet~al.}{2018}]%
        {preventingnn18}
\bibfield{author}{\bibinfo{person}{Mihailo Isakovet~al.}}
  \bibinfo{year}{2018}\natexlab{}.
\newblock \showarticletitle{Preventing Neural Network Model Exfiltration in
  Machine Learning Hardware Accelerators}. In
  \bibinfo{booktitle}{\emph{AsianHOST '18}}.
\newblock


\bibitem[\protect\citeauthoryear{Juuti~et al.}{Juuti~et al.}{2019}]%
        {juuti2019prada}
\bibfield{author}{\bibinfo{person}{Mika Juuti~et al.}}
  \bibinfo{year}{2019}\natexlab{}.
\newblock \showarticletitle{{PRADA}: Protecting Against {DNN} Model Stealing
  Attacks}. In \bibinfo{booktitle}{\emph{EuroS\&P '19}}.
\newblock


\bibitem[\protect\citeauthoryear{Juvekar~et al.}{Juvekar~et al.}{2018}]%
        {gazelle18}
\bibfield{author}{\bibinfo{person}{Chiraag Juvekar~et al.}}
  \bibinfo{year}{2018}\natexlab{}.
\newblock \showarticletitle{{GAZELLE}: A Low Latency Framework for Secure
  Neural Network Inference}. In \bibinfo{booktitle}{\emph{{USENIX} Security
  '18}}.
\newblock


\bibitem[\protect\citeauthoryear{Kocher~et al.}{Kocher~et al.}{2011}]%
        {kocher2011introduction}
\bibfield{author}{\bibinfo{person}{Paul Kocher~et al.}}
  \bibinfo{year}{2011}\natexlab{}.
\newblock \showarticletitle{Introduction to Differential Power Analysis}.
\newblock \bibinfo{journal}{\emph{Journal of Cryptographic Engineering}}
  \bibinfo{volume}{1} (\bibinfo{year}{2011}).
\newblock


\bibitem[\protect\citeauthoryear{Kogge~et al.}{Kogge~et al.}{1973}]%
        {koggestone}
\bibfield{author}{\bibinfo{person}{Peter~M Kogge~et al.}}
  \bibinfo{year}{1973}\natexlab{}.
\newblock \showarticletitle{A Parallel Algorithm for the Efficient Solution of
  a General Class of Recurrence Equations}.
\newblock \bibinfo{journal}{\emph{IEEE Trans. Comput.}} \bibinfo{volume}{100},
  \bibinfo{number}{8} (\bibinfo{year}{1973}).
\newblock


\bibitem[\protect\citeauthoryear{Lehman~et al.}{Lehman~et al.}{1961}]%
        {csa}
\bibfield{author}{\bibinfo{person}{M Lehman~et al.}}
  \bibinfo{year}{1961}\natexlab{}.
\newblock \showarticletitle{Skip Techniques for High-Speed Carry-Propagation in
  Binary Arithmetic Units}.
\newblock \bibinfo{journal}{\emph{IRE Transactions on Electronic Computers}}
  \bibinfo{number}{4} (\bibinfo{year}{1961}).
\newblock


\bibitem[\protect\citeauthoryear{Lowd~et al.}{Lowd~et al.}{2005}]%
        {advlearn05}
\bibfield{author}{\bibinfo{person}{Daniel Lowd~et al.}}
  \bibinfo{year}{2005}\natexlab{}.
\newblock \showarticletitle{Adversarial Learning}. In
  \bibinfo{booktitle}{\emph{Proceedings of the Eleventh ACM SIGKDD
  International Conference on Knowledge Discovery in Data Mining}}.
\newblock


\bibitem[\protect\citeauthoryear{Mangard~et al.}{Mangard~et al.}{2005}]%
        {maskedcmosleak}
\bibfield{author}{\bibinfo{person}{Stefan Mangard~et al.}}
  \bibinfo{year}{2005}\natexlab{}.
\newblock \showarticletitle{Side-Channel Leakage of Masked CMOS Gates}. In
  \bibinfo{booktitle}{\emph{Topics in Cryptology -- CT-RSA 2005}}.
\newblock


\bibitem[\protect\citeauthoryear{Mangard~et al.}{Mangard~et al.}{2008}]%
        {mangard2008power}
\bibfield{author}{\bibinfo{person}{Stefan Mangard~et al.}}
  \bibinfo{year}{2008}\natexlab{}.
\newblock \bibinfo{booktitle}{\emph{Power analysis attacks: Revealing the
  secrets of smart cards}}. Vol.~\bibinfo{volume}{31}.
\newblock \bibinfo{publisher}{Springer Science \& Business Media}.
\newblock


\bibitem[\protect\citeauthoryear{Mishra~et al.}{Mishra~et al.}{2020}]%
        {delphi20}
\bibfield{author}{\bibinfo{person}{Pratyush Mishra~et al.}}
  \bibinfo{year}{2020}\natexlab{}.
\newblock \showarticletitle{DELPHI: A {Cryptographic} {Inference} {Service} for
  {Neural} {Networks}}. In \bibinfo{booktitle}{\emph{USENIX Security '20}}.
\newblock


\bibitem[\protect\citeauthoryear{Moos~et al.}{Moos~et al.}{2017}]%
        {moos2017static}
\bibfield{author}{\bibinfo{person}{Thorben Moos~et al.}}
  \bibinfo{year}{2017}\natexlab{}.
\newblock \showarticletitle{Static Power Side-Channel Analysis of a Threshold
  Implementation Prototype Chip}. In \bibinfo{booktitle}{\emph{DATE '17}}.
\newblock


\bibitem[\protect\citeauthoryear{Nassar~et al.}{Nassar~et al.}{2010}]%
        {nassar2010bcdl}
\bibfield{author}{\bibinfo{person}{Maxime Nassar~et al.}}
  \bibinfo{year}{2010}\natexlab{}.
\newblock \showarticletitle{{BCDL}: A High Speed Balanced {DPL} for {FPGA} with
  Global Precharge and No Early Evaluation}. In \bibinfo{booktitle}{\emph{DATE
  '10}}.
\newblock


\bibitem[\protect\citeauthoryear{NIST}{NIST}{2019}]%
        {mlNIST}
\bibfield{author}{\bibinfo{person}{NIST}.} \bibinfo{year}{2019}\natexlab{}.
\newblock \bibinfo{booktitle}{\emph{A Taxonomy and Terminology of Adversarial
  Machine Learning}}.
\newblock
\urldef\tempurl%
\url{https://nvlpubs.nist.gov/nistpubs/ir/2019/NIST.IR.8269-draft.pdf}
\showURL{%
\tempurl}


\bibitem[\protect\citeauthoryear{Oh~et al.}{Oh~et al.}{2019}]%
        {oh2019towards}
\bibfield{author}{\bibinfo{person}{Seong~Joon Oh~et al.}}
  \bibinfo{year}{2019}\natexlab{}.
\newblock \showarticletitle{Towards Reverse-Engineering Black-Box Neural
  Networks}.
\newblock In \bibinfo{booktitle}{\emph{Explainable AI: Interpreting, Explaining
  and Visualizing Deep Learning}}.
\newblock


\bibitem[\protect\citeauthoryear{Oswald~et al.}{Oswald~et al.}{2005}]%
        {OMPR05}
\bibfield{author}{\bibinfo{person}{Elisabeth Oswald~et al.}}
  \bibinfo{year}{2005}\natexlab{}.
\newblock \showarticletitle{A Side-Channel Analysis Resistant Description of
  the {AES} S-Box}. In \bibinfo{booktitle}{\emph{Fast Software Encryption}}.
\newblock


\bibitem[\protect\citeauthoryear{Rastegari~et al.}{Rastegari~et al.}{2016}]%
        {rastegari2016xnor}
\bibfield{author}{\bibinfo{person}{Mohammad Rastegari~et al.}}
  \bibinfo{year}{2016}\natexlab{}.
\newblock \showarticletitle{XNOR-Net: Imagenet Classification using Binary
  Convolutional Neural Networks}. In \bibinfo{booktitle}{\emph{ECCV '16}}.
\newblock


\bibitem[\protect\citeauthoryear{Reith~et al.}{Reith~et al.}{2019}]%
        {RST19}
\bibfield{author}{\bibinfo{person}{Robert~Nikolai Reith~et al.}}
  \bibinfo{year}{2019}\natexlab{}.
\newblock \showarticletitle{Efficiently Stealing Your Machine Learning Models}.
  In \bibinfo{booktitle}{\emph{18th ACM Workshop on Privacy in the Electronic
  Society}}.
\newblock


\bibitem[\protect\citeauthoryear{Reparaz~et al.}{Reparaz~et al.}{2016}]%
        {reparaz16-2}
\bibfield{author}{\bibinfo{person}{Oscar Reparaz~et al.}}
  \bibinfo{year}{2016}\natexlab{}.
\newblock \showarticletitle{Additively Homomorphic Ring-{LWE} Masking}. In
  \bibinfo{booktitle}{\emph{International Workshop on Post-Quantum
  Cryptography}}.
\newblock


\bibitem[\protect\citeauthoryear{Riazi~et al.}{Riazi~et al.}{2019}]%
        {xonn19}
\bibfield{author}{\bibinfo{person}{M~Sadegh Riazi~et al.}}
  \bibinfo{year}{2019}\natexlab{}.
\newblock \showarticletitle{{XONN}: {XNOR}-based Oblivious Deep Neural Network
  Inference}. In \bibinfo{booktitle}{\emph{{USENIX} Security '19}}.
\newblock


\bibitem[\protect\citeauthoryear{Riscure}{Riscure}{2019}]%
        {inspector}
\bibfield{author}{\bibinfo{person}{Riscure}.} \bibinfo{year}{2019}\natexlab{}.
\newblock \bibinfo{booktitle}{\emph{Riscure Inspector}}.
\newblock
\urldef\tempurl%
\url{https://www.riscure.com/uploads/2017/08/inspector_brochure.pdf}
\showURL{%
Retrieved May 7, 2020 from \tempurl}


\bibitem[\protect\citeauthoryear{Rouhani~et al.}{Rouhani~et al.}{2018}]%
        {rouhani2018deepsigns}
\bibfield{author}{\bibinfo{person}{Bita~Darvish Rouhani~et al.}}
  \bibinfo{year}{2018}\natexlab{}.
\newblock \showarticletitle{Deepsigns: A Generic Watermarking Framework for IP
  Protection of Deep Learning Models}.
\newblock \bibinfo{journal}{\emph{arXiv preprint arXiv:1804.00750}}
  (\bibinfo{year}{2018}).
\newblock


\bibitem[\protect\citeauthoryear{Schneider~et al.}{Schneider~et al.}{2015}]%
        {boma-adder15}
\bibfield{author}{\bibinfo{person}{Tobias Schneider~et al.}}
  \bibinfo{year}{2015}\natexlab{}.
\newblock \showarticletitle{Arithmetic Addition over Boolean Masking}. In
  \bibinfo{booktitle}{\emph{Applied Cryptography and Network Security}}.
\newblock


\bibitem[\protect\citeauthoryear{Schneider~et al.}{Schneider~et al.}{2016}]%
        {schneider2016leakage}
\bibfield{author}{\bibinfo{person}{Tobias Schneider~et al.}}
  \bibinfo{year}{2016}\natexlab{}.
\newblock \showarticletitle{Leakage Assessment Methodology}.
\newblock \bibinfo{journal}{\emph{Journal of Cryptographic Engineering}}
  \bibinfo{volume}{6}, \bibinfo{number}{2} (\bibinfo{year}{2016}),
  \bibinfo{pages}{85--99}.
\newblock


\bibitem[\protect\citeauthoryear{Strubell~et al.}{Strubell~et al.}{2019}]%
        {strubell2019energy}
\bibfield{author}{\bibinfo{person}{Emma Strubell~et al.}}
  \bibinfo{year}{2019}\natexlab{}.
\newblock \showarticletitle{Energy and Policy Considerations for Deep Learning
  in NLP}.
\newblock \bibinfo{journal}{\emph{arXiv preprint arXiv:1906.02243}}
  (\bibinfo{year}{2019}).
\newblock


\bibitem[\protect\citeauthoryear{Technology}{Technology}{2020}]%
        {picoscope}
\bibfield{author}{\bibinfo{person}{Pico Technology}.}
  \bibinfo{year}{2020}\natexlab{}.
\newblock \bibinfo{booktitle}{}.
\newblock
\urldef\tempurl%
\url{https://www.picotech.com/oscilloscope/3000/picoscope-3000-oscilloscope-specifications}
\showURL{%
\tempurl}


\bibitem[\protect\citeauthoryear{Tiri~et al.}{Tiri~et al.}{2004}]%
        {tiri2004logic}
\bibfield{author}{\bibinfo{person}{Kris Tiri~et al.}}
  \bibinfo{year}{2004}\natexlab{}.
\newblock \showarticletitle{A Logic Level Design Methodology for a Secure {DPA}
  Resistant {ASIC} or {FPGA} implementation}. In \bibinfo{booktitle}{\emph{DATE
  '04}}, Vol.~\bibinfo{volume}{1}.
\newblock


\bibitem[\protect\citeauthoryear{Tramer~et a;.}{Tramer~et a;.}{2018}]%
        {tramer2018slalom}
\bibfield{author}{\bibinfo{person}{Florian Tramer~et a;.}}
  \bibinfo{year}{2018}\natexlab{}.
\newblock \showarticletitle{Slalom: Fast, verifiable and private execution of
  neural networks in trusted hardware}.
\newblock \bibinfo{journal}{\emph{arXiv preprint arXiv:1806.03287}}
  (\bibinfo{year}{2018}).
\newblock


\bibitem[\protect\citeauthoryear{Tram{\`e}r~et al.}{Tram{\`e}r~et al.}{2016}]%
        {tramer2016stealing}
\bibfield{author}{\bibinfo{person}{Florian Tram{\`e}r~et al.}}
  \bibinfo{year}{2016}\natexlab{}.
\newblock \showarticletitle{Stealing Machine Learning Models via Prediction
  APIs}. In \bibinfo{booktitle}{\emph{USENIX Security '16}}.
\newblock


\bibitem[\protect\citeauthoryear{Trichina~et al.}{Trichina~et al.}{2004}]%
        {TKL05}
\bibfield{author}{\bibinfo{person}{Elena Trichina~et al.}}
  \bibinfo{year}{2004}\natexlab{}.
\newblock \showarticletitle{Small Size, Low Power, Side Channel-Immune AES
  Coprocessor: Design and Synthesis Results}. In \bibinfo{booktitle}{\emph{AES
  '04}}.
\newblock


\bibitem[\protect\citeauthoryear{Umuroglu~et al.}{Umuroglu~et al.}{2017}]%
        {umuroglu2017finn}
\bibfield{author}{\bibinfo{person}{Yaman Umuroglu~et al.}}
  \bibinfo{year}{2017}\natexlab{}.
\newblock \showarticletitle{FINN: A Framework for Fast, Scalable Binarized
  Neural Network Inference}. In \bibinfo{booktitle}{\emph{FPGA '17}}.
\newblock


\bibitem[\protect\citeauthoryear{Wei~et al.}{Wei~et al.}{2018}]%
        {wei2018know}
\bibfield{author}{\bibinfo{person}{Lingxiao Wei~et al.}}
  \bibinfo{year}{2018}\natexlab{}.
\newblock \showarticletitle{I Know What You See: Power Side-Channel Attack on
  Convolutional Neural Network Accelerators}. In
  \bibinfo{booktitle}{\emph{ACSAC '18}}.
\newblock


\bibitem[\protect\citeauthoryear{Xiang~et al.}{Xiang~et al.}{2020}]%
        {xiang2020open}
\bibfield{author}{\bibinfo{person}{Yun Xiang~et al.}}
  \bibinfo{year}{2020}\natexlab{}.
\newblock \showarticletitle{Open {DNN} Box by Power Side-Channel Attack}.
\newblock \bibinfo{journal}{\emph{IEEE Transactions on Circuits and Systems II:
  Express Briefs}} (\bibinfo{year}{2020}).
\newblock


\bibitem[\protect\citeauthoryear{Xilinx}{Xilinx}{2010}]%
        {ugSpartan6}
\bibfield{author}{\bibinfo{person}{Xilinx}.} \bibinfo{year}{2010}\natexlab{}.
\newblock \bibinfo{booktitle}{\emph{Spartan-6 FPGA Configurable Logic Block
  User Guide}}.
\newblock
\urldef\tempurl%
\url{https://www.xilinx.com/support/documentation/user_guides/ug384.pdf}
\showURL{%
\tempurl}


\bibitem[\protect\citeauthoryear{Yan~et al.}{Yan~et al.}{2020}]%
        {yan2018cache}
\bibfield{author}{\bibinfo{person}{Mengjia Yan~et al.}}
  \bibinfo{year}{2020}\natexlab{}.
\newblock \showarticletitle{Cache Telepathy: Leveraging Shared Resource Attacks
  to Learn {DNN} Architectures}. In \bibinfo{booktitle}{\emph{{USENIX} Security
  '20}}.
\newblock


\bibitem[\protect\citeauthoryear{Yu~et al.}{Yu~et al.}{2020}]%
        {yudeepem}
\bibfield{author}{\bibinfo{person}{Honggang Yu~et al.}}
  \bibinfo{year}{2020}\natexlab{}.
\newblock \showarticletitle{DeepEM: Deep Neural Networks Model Recovery through
  EM Side-Channel Information Leakage}.
\newblock  (\bibinfo{year}{2020}).
\newblock


\bibitem[\protect\citeauthoryear{Yu~et al.}{Yu~et al.}{2007}]%
        {yu2007secure}
\bibfield{author}{\bibinfo{person}{Pengyuan Yu~et al.}}
  \bibinfo{year}{2007}\natexlab{}.
\newblock \showarticletitle{Secure {FPGA} Circuits using Controlled Placement
  and Routing}. In \bibinfo{booktitle}{\emph{(CODES+ISSS) '07}}.
\newblock


\bibitem[\protect\citeauthoryear{Zhao~et al.}{Zhao~et al.}{2018}]%
        {zhao2018fpga}
\bibfield{author}{\bibinfo{person}{Mark Zhao~et al.}}
  \bibinfo{year}{2018}\natexlab{}.
\newblock \showarticletitle{FPGA-based Remote Power Side-Channel Attacks}. In
  \bibinfo{booktitle}{\emph{2018 IEEE Symposium on Security and Privacy (SP)}}.
\newblock


\end{thebibliography}

\end{document}